\documentclass[prologue, dvipsnames, sigplan, screen]{acmart}

\usepackage{ragged2e}
\usepackage{caption}
\usepackage{subcaption}
\usepackage[utf8]{inputenc}
\usepackage{tabularx}
\usepackage{listings}
\usepackage{amsmath}
\usepackage{soul} 
\usepackage{xspace}
\usepackage{float}
\usepackage{enumitem}
\usepackage{microtype}
\usepackage[T1]{fontenc}

\newcommand{\g}{Glenside\xspace}

\newcommand{\accesspatternshape}[2]{$($$\left( #1 \right)$, $\left( #2 \right)$$)$}
\newcommand{\itc}{\texttt{im2col}\xspace}
\newcommand{\ctd}{\texttt{conv2d}\xspace}

\newcommand{\tcd}[1]{\texttt{#1}}
\newcommand{\mcd}[1]{\mathrm{\tcd{#1}}}

\title[\g]{Pure Tensor Program Rewriting via Access Patterns}

\subtitle{(Representation Pearl)}

\author{
  Gus Henry Smith,
  Andrew Liu,
  Steven Lyubomirsky,
  Scott Davidson,\\
  Joseph McMahan,
  Michael Taylor,
  Luis Ceze,
  Zachary Tatlock}
\email{{
    gussmith,
    andy99,
    sslyu,
    stdavids,
    jmcmahan,
    profmbt,
    luisceze,
    ztatlock
  }@cs.washington.edu}
\affiliation{%
  \institution{Paul G.~Allen School of Computer Science \& Engineering at the University of Washington}
  \streetaddress{Some street address}
  \city{Seattle}
  \state{WA}
  \country{USA}
  \postcode{some post code}
}

\setcopyright{acmlicensed}\acmConference[MAPS '21]{Proceedings of the 5th ACM SIGPLAN International Symposium on Machine Programming}{June 21, 2021}{Virtual Event, Canada}
\acmPrice{15.00}
\acmDOI{10.1145/3460945.3464953}
\acmYear{2021}
\copyrightyear{2021}
\acmSubmissionID{pldiws21mapsmain-p37-p}
\acmISBN{978-1-4503-8467-4/21/06}
\acmConference[MAPS '21]{Proceedings of the 5th ACM SIGPLAN International Symposium on Machine Programming}{June 21, 2021}{Virtual, Canada}
\acmBooktitle{Proceedings of the 5th ACM SIGPLAN International Symposium on Machine Programming (MAPS '21), June 21, 2021, Virtual, Canada}


\begin{abstract}

Tensor kernels in machine learning (ML)
  often correspond to pure mathematical expressions,
  making term rewriting an attractive strategy
  for optimization and mapping to specialized hardware accelerators.
However,
  existing ML intermediate representations (IRs)
  tend to either be \textit{pure but high-level},
  making low-level rewrites
  to hardware targets inexpressible,
  or \textit{low-level but impure},
  hampering the use of term rewriting altogether.

This paper introduces \g,
  a pure IR whose core abstraction---%
  the \textit{access pattern}---%
  enables
  low-level,
  layout-aware,
  hardware-centric
  program rewrites.
We demonstrate how term rewriting
  in \g
  can be used to 
  map program fragments
  to hardware accelerator invocations
  and
  automatically discover
  classic data layout transformations
  like \tcd{im2col}.
\g establishes a new foundation for
  exploring further term rewriting techniques
  in optimizing low-level tensor programs.

\end{abstract}

\begin{document}

\maketitle

\section{Introduction}
    
\begin{figure}
    \centering
    \includegraphics[width=.7\linewidth]{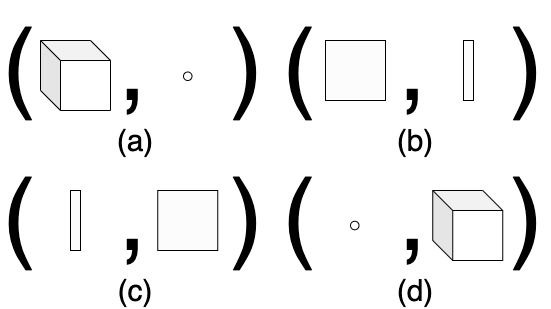}
    \caption{
      Four access patterns,
        representing different ways
        a
        tensor program
        (or \textit{kernel})
        might access
        the same 3D tensor. 
      For example, (c) represents
        accessing a 3D tensor as
        a vector of 2D matrices.}
    \label{fig:access-pattern-examples}
    \vspace{-1em}
\end{figure}


Machine learning (ML) and other
  high-performance computing (HPC)
  applications increasingly rely on
  specialized hardware accelerators to
  provide speed and energy efficiency~\cite{jouppi2017tpu, krizhevsky2012conv, reuther2019survey}.
This trend has highlighted the need
  for flexible accelerator support
  in domain-specific compilers like
  Halide~\cite{halide},
  TVM~\cite{tvm},
  TensorFlow/MLIR~\cite{tensorflow, mlir}, and
  PyTorch~\cite{pytorch}.

Adding accelerator support to
  an existing compiler typically
  uses custom pattern matching to
  map expensive tensor operations
  from applications down to
  accelerator invocations~\cite{
    yang2020interstellar, byoc}.
Pattern matching often additionally relies on
  various other transformations
  to canonicalize intermediate representations (IRs)
  and massage data layouts into
  formats matching accelerator requirements~\cite{nvidia2020nhwc}.
Even with these changes,
  users may need to manually modify their application to
  help the compiler discover opportunities
  for dispatching operations to accelerators, 
  such as by changing data types or unrolling loops.
    
In principle, term rewriting techniques~\cite{baader1998term}
  should be able to facilitate many of
  these transformation and mapping tasks
  within a compiler.
Halide and TVM already rely
  on extensive rewrite systems for
  optimizing scalar computations and
  simplifying loop bounds in order to
  support further downstream optimizations~\cite{newcomb2020halide-rewrite,
  hagedorn2020func-high-perf}.

Unfortunately, existing IRs in compilers for
  array/tensor programming DSLs tend to
  present abstraction and granularity mismatches
  that hamper term rewriting approaches.
Term rewriting is most easily applied in
  \textit{pure} (side effect--free) IRs
  that support equational reasoning.
At the same time,
  mapping to accelerators requires considering
  low-level hardware details like data layout.
Existing pure IRs for ML frameworks are used
  primarily for high-level transformations
  (e.g., type elaboration and inlining)
  and do not expose low-level data layout details~\cite{
    relay}.
On the other hand,
  IRs used for crucial lower-level optimizations like
  operator fusion must support
  precise reasoning about memory use,
  and therefore are typically impure,
  hampering term rewriting.

To help mitigate such impedance mismatches,
  we present \textit{\g},\footnote{Publicly available at \url{https://github.com/gussmith23/glenside}.}
  a pure tensor program IR
  that enables hardware-level term rewriting.
\g is based on a simple
  \textit{access pattern} abstraction that
  supports expressing and reasoning about
  data layout transformations via
  syntactic rewrite rules.
When combined with standard arithmetic rewrites
  for per-tensor-element computations,
  access patterns enable implementing complex
  transformations for accelerator support as
  compositions of simple rewrites.

Tensors are traditionally characterized
  by their \textit{shape},
  an $n$-tuple 
  of positive integers
  indicating the size of each
  of a tensor's dimensions.
Access patterns instead characterize
  each tensor with two shapes, e.g.,
  \accesspatternshape{x}{y, z}, separating
  the dimensions which are \textit{iterated over} from
  the dimensions which are \textit{computed on.}
Figure~\ref{fig:access-pattern-examples}(c)
  depicts an example where a 3D tensor's
  first dimension is iterated over and
  some computation applied to each
  corresponding 2D matrix.

We demonstrate how \g
  enables implementing representative
  hardware-level transformation via term rewriting,
  including mapping computations
  to systolic arrays~\cite{jouppi2017tpu}
  (a common hardware module in ML accelerators)
  and automatically discovering the
  \tcd{im2col} data layout transformation~\cite{im2col},
  which enables mapping 2D convolutions
  to matrix multiplication hardware.
In particular,
  by employing \textit{equality saturation}~\cite{willsey2021egg},
  these transformations ``fall out for free''
  (i.e., without any carefully crafted
  rewrite orderings~\cite{phase-ordering}),
  from a handful of general rewrites concerning tensor
  transposition, Cartesian product, dot product, etc.,
  expressed in terms of access patterns.

To summarize, our contributions include:
\begin{itemize}
\item \textit{Access patterns},
  a tensor representation that employs a
  simple, extended tensor shape type to
  distinguish iteration and computation dimensions

\item The \g IR,
  a pure compiler IR that facilitates 
  term rewriting to enable support for
  specialized accelerators
  
\item A library of generic rewrites over \g programs
  
\item Case studies demonstrating how
  \g enables automatically discovering
  key transformations for mapping
  applications to custom accelerators
  via equality saturation with the
  \tcd{egg}~\cite{willsey2021egg} library.
\end{itemize}

The rest of the paper is organized as follows:
Section~\ref{sec:background} provides background
  and briefly surveys closely related work.
Section~\ref{sec:matmul} motivates
  \g via a running example exploring
  pure matrix multiplication.
Section~\ref{sec:glenside} details the
  design and implementation of \g.
Section~\ref{sec:case-studies} details
  case studies showing the potential
  benefits of \g's term rewriting
  approach to low-level tensor program
  transformations.

\section{Background and Related Work}
\label{sec:background}

\g is designed to help target
  tensor hardware accelerators and
  builds on past work in
  tensor IRs and term rewriting.


\subsection{Machine Learning Accelerators}

A variety of accelerators~\cite{
    jouppi2017tpu, chen2016eyeriss, moreau2018vta, markidis2018tensorcore, nvdla}
  have been developed 
  to provide efficient implementations
  of tensor operators for ML applications.
These devices accelerate tensor operators 
  through hardware parallelism, 
  simultaneously applying related operations
  across many tensors in the accelerator's memory (which are often laid out according to custom rules that facilitate hardware optimization).
Tensor program compilers must translate
  expensive application code fragments
  down to accelerator invocations that
  adhere to these layout rules,
  which often involves both
  (a) higher-level transformations like
  tensor reshaping to match accelerator size bounds and
  loop unrolling to expose optimization opportunities, and
  (b) lower-level transformations like
  operator fusion and \tcd{im2col}
  to match accelerator calling conventions and
  even implement different operations
  using the same accelerator,
  e.g., on systolic arrays~\cite{im2col, jia2014semantic}.
  

\subsection{Tensor IRs and Compilers}



Tensor compilers for ML and HPC applications strive
  to balance clear, high-level operator semantics
  and support for the low-level optimizations
  necessary to target specialized accelerators.
Halide~\cite{ragan2013halide}
  achieves this balance by separating
  operator \textit{specifications} (what is computed) from
  \textit{schedules} (how, when, and where
  each output element is generated).
This style of separation has proven
  highly effective across both
  application domains and hardware targets;
  numerous compilers including TVM~\cite{chen2018tvm},
  FireIron~\cite{hagedorn2020fireiron},
  LIFT~\cite{lift}, and Accelerate~\cite{accelerate}
  follow variations of this strategy.
  
The specification/schedule separation approach
  allows the same high-level program (specification)
  to be flexibly optimized for and mapped to
  different hardware targets by applying different schedules.
From this perspective,
  schedules represent different rewriting strategies
  to explore various loop ordering and memory layouts;
  in LIFT and Accelerate these
  take the form of functional combinators
  closely related to \g's approach.
As in classic term rewriting,
  experts must often carefully craft
  schedules for each target to achieve
  the best performance and mitigate
  phase ordering challenges~\cite{phase-ordering},
  though recent projects have produced promising results
  in automatic scheduling~\cite{
    chen2018autotvm, zheng2020ansor, anderson2020learning}.

Other tensor IRs like
  TACO~\cite{taco}, Keops~\cite{keops},
  and COMET~\cite{tian2021highperformance}
  rely on \textit{index notation}\footnote{
    Index notation is closely related to
    ``Einstein notation'' where reduction
    indices are implicit.}
  to concisely express tensor operators
  and simplify optimization by
  uniformly representing
  per-output-element computations.
These approaches also rely on
  rewriting passes to generate
  kernel implementations specialized to
  tensor sparsity / density,
  operator combinations arising in
  the source application, and
  details of the target hardware.
In Section~\ref{sec:matmul} we discuss
  some of the tradeoffs of these approaches
  with respect to other rewriting strategies.
 
Finally, polyhedral compilers~\cite{polyhedral-survey}
  like Tensor Comprehensions~\cite{vasilache2018tensor}
  and Tiramisu~\cite{tiramisu}
  optimize loop-filled programs
  by modeling loop nests as polyhedra
  and applying geometric transformations.
The polyhedral approach exploits
  regular loop structure,
  but is also restricted
  to geometrically affine transformations.
In contrast, term rewriting is
  neither guided nor restricted by
  geometric constraints, making
  these approaches broadly complementary.

\subsection{Term Rewriting and Equality Saturation}

Term rewriting is a classic
  program optimization technique~\cite{baader1998term}
  that relies on iteratively applying
  rewrite rules of the form $\ell \xrightarrow{} r$:
  when part of a program
  matches the pattern $\ell$
  under substitution $\sigma$,
  it is rewritten into $\sigma(r)$.
This approach is ubiquitous,
  frequently used in both mainstream and DSL compilers
  to implement features including preprocessing,
  strength reductions, and
  peephole optimizations~\cite{garavel2018rewrite-context}.
  
Classic term rewriting systems where
  rewrites are applied destructively suffer
  phase ordering problems~\cite{phase-ordering}:
  the order in which rewrites are applied can
  enhance or severely diminish performance.
Recent work has shown how program synthesis
  can help address this challenge in
  peephole optimizers like Halide's
  scalar expression rewriter~\cite{
    newcomb2020halide-rewrite}.
  
Advances in alternate rewriting techniques
  like equality saturation~\cite{tate2009equality}
  also mitigate phase ordering by exploiting
  the e-graph data structure from SMT solvers
  to repeatedly apply all rewrites simultaneously,
  thus obviating rule ordering considerations.
In particular,
  the \tcd{egg} library~\cite{willsey2021egg}
  has been used to develop new synthesis and
  optimization tools across diverse domains~\cite{
    herbie, szalinski, wang2020spores},
  including DSP compiler vectorization~\cite{
    vanhattum2021vectorization} and
  tensor computation graphs~\cite{yang2021equality}.

\g provides the first tensor IR
  amenable to equality saturation by
  introducing access patterns to
  provide pure, higher order tensor
  kernel combinators that support
  rank-polymorphism without the need
  for binding structures like
  anonymous functions or index notation.

\section{From Pure \texttt{matMul} to IR Design Goals}
\label{sec:matmul}

  
Applying functional programming techniques
  and term rewriting to tensor IRs
  requires careful design.
For example,
  we must ensure that operators be compositional
  with respect to tensor shapes
  and that the representation support
  generic rules within the
  target rewrite engine.
To highlight such constraints and
  motivate access patterns in \g,
  this section illustrates potential pitfalls
  with a simple matrix multiplication example.

\subsection{Pure Matrix Multiplication}
\label{subsec:pure-matmul}

We write
  \tcd{f64} for the type of 64-bit floats and
  \tcd{[A]} for vectors over type \tcd{A}.
Using this notation, we can specify operators like
  dot product and 2D matrix transpose as:
\begin{align*}
    \mcd{dotProd} &
    \mcd{ : [f64] * [f64] -> f64} \\
    \mcd{trans2} &
    \mcd{ : [[f64]] -> [[f64]]}
\end{align*} 






\noindent
Implementing 2D matrix multiplication
  on inputs $P$ and $Q$ requires computing
  an output matrix $R$ where
  $R_{ij} = \Sigma_k P_{ik} \, Q_{kj}
          =  P_i \cdot Q^{T}_{j}$. 
The need to compute \tcd{dotProd} for every pair
  of a row from $P$ and a column from $Q$
  suggests map and Cartesian product operators
  which we might specify with:
\begin{align*}
    \mcd{map} &
    \mcd{ : (A -> B) * [A] -> [B]} \\
    \mcd{cartProd} &
    \mcd{ : [A] * [B] -> [A * B]}
\end{align*}
Naively, we can almost implement matrix multiplication as:
{\color{red} \begin{align*}
  & \mcd{matMul(P, Q) :=} \\
  & \;\;\;\;\; \mcd{map(dotProd, cartProd(P, trans2(Q)))}
\end{align*} }
However, the result type will have been
  flattened to just {\color{red}\tcd{[f64]}},
  making it impossible to compose with other matrix
  operators that expect \tcd{[[f64]]} inputs.

Our first problem is that
  the \tcd{cartProd} specification above
  ``forgets'' the shape of its arguments.
We could change this specification to
  arrange the output as a matrix:
$$
  \mcd{cartProd2D : [A] * [B] -> [[A * B]]}
$$
But this result type prevents
  directly mapping \tcd{dotProd}.\footnote{
    This simple type does not specify how
    \tcd{cartProd2D} orders its output
    relative to its input vectors.
    We assume the order
    expected for matrix multiplication.}
Now the problem is that \tcd{map}
  only applies a computation by iterating
  over the first (outermost) dimension of a tensor.
If we specialize \tcd{map} to iterate
  over the second dimension:
$$
  \mcd{mapAt2 : (A -> B) * [[A]] -> [[B]]}
$$
then we can implement a compositional
  \tcd{matMul} operator that correctly produces
  results of type \tcd{[[f64]]} as:
\begin{align*}
  & \mcd{matMul(P, Q) :=} \\
  & \;\;\;\;\; \mcd{mapAt2(dotProd, cartProd2D(P, trans2(Q)))}
\end{align*}

\subsection{\g Design Constraints and Goals}

This style of pure, higher-order functional
  program representation enables
  term rewriting and equational reasoning
  via rules like:
\begin{align*}
  \mcd{dotProd(P, Q)}
    & \leftrightsquigarrow
      \mcd{dotProd(Q, P)} \\[2pt]
  \mcd{trans2(trans2(P))}
    & \leftrightsquigarrow
      P \\[2pt]
  \mcd{map(f, map(g, P))}
    & \leftrightsquigarrow
      \mcd{map(f$\,\circ\,$g, P)} \\[2pt]
  \mcd{mapAt2(f, trans2(P))}
    & \leftrightsquigarrow
      \mcd{trans2(mapAt2(f, P))} 
\end{align*}


However, some of these rules depend on the
  shapes of dimension-specific operators aligning.
What happens when we need to support
  higher-dimensional tensors?
Without a mechanism to abstract
  which dimensions of a tensor
  are being iterated as opposed to computed over,
  we would have to generate versions of
  each rule for every combination of dimensions.
Worse, these problems
  do not only affect rewrite rules;
  they also lead to code blowup just to
  specify all the variants of tensor kernels
  that arise in practice.

One strategy to address these challenges is
  adding support for anonymous functions (``lambdas''),
  currying, and closures to the 
  tensor program representation.
These features can provide sufficient
  flexibility to handle shape alignment
  issues that otherwise may require
  dimension-specific operators like
  \tcd{cartProd2D} and \tcd{mapAt2} above.
For example, given curried versions
  of \tcd{dotProd} and \tcd{map},
  we could have used such features
  to implement a curried \tcd{matMul} as:
\begin{align*}
  & \mcd{matMul' P Q :=} \\
  & \;\; \mcd{ \
      map' ($\boldsymbol\lambda\,$r =>\
        map' (dotProd' r) (trans2 Q)) P}
\end{align*}
Alternatively, some IRs rely on index notation
  for even pithier implementations like:
$$
  \mcd{matMul(P,Q)[i,j] := dotProd(P[i], trans2(Q)[j])}
$$

Unfortunately, these approaches all rely on some
  form of \textit{name binding} which can
  significantly complicate term rewriting.
Rewriting under binders,
  whether explicitly in the form of lambdas
  or implicitly with index notation,
  requires additionally analyzing the
  potential \textit{contexts}
  (what names are bound to)
  of every subexpression.
While it is still technically possible to
  apply state-of-the-art rewrite engines
  like \tcd{egg}~\cite{willsey2021egg}
  via explicit variable substitution rules and
  free variable analyses,
  we have found the additional complexity
  and rewrite search space blow up
  substantially eliminate the potential advantages
  of term rewriting in such IR designs.

All the above constraints inform \g's key design goal:
  providing an IR that flexibly supports specifying and
  composing higher-order tensor operators\footnote{
    As \tcd{map} and \tcd{mapAt2} in 
    Section~\ref{subsec:pure-matmul} illustrate,
    an IR can support higher-order operators without
    necessarily providing lambdas, currying, or closures.}
  over arbitrary dimensions while still enabling
  high-performance term rewriting techniques
  like equality saturation.
In the rest of this paper,
  we show how \textit{access patterns} enable achieving
  these goals with a focus on applications to
  mapping application fragments down to
  specialized hardware accelerators.

\section{\g}
\label{sec:glenside}
  
\begin{table*}
    \centering
    \caption{\g's access pattern transformers.}
    \label{tab:access-pattern-transformers}
    \begin{tabularx}{\linewidth}{lXX}
    Transformer 
    & Input(s)
    & Output Shape  \\
    \hline
    
    \texttt{access} 
    &
    \accesspatternshape{a_0,\dots}{\dots, a_n}
    and non-negative integer $i$
    & 
  \accesspatternshape
  {a_0, \dots, a_{i-1}}{a_i,\dots, a_n}
    \\
    
    \texttt{transpose} &
    \accesspatternshape{a_0,\dots}{\dots, a_n},  $\ell$ (a permutation of $(0, \dots, n-1)$) &
    \accesspatternshape{a_{\ell_0},\dots}{\dots, a_{\ell_n}}
    \\
    
    \texttt{cartProd} 
    &
    \accesspatternshape{a_0,\dots, a_n}{c_0, \dots, c_p},  \accesspatternshape{b_0,\dots, b_m}{c_0, \dots, c_p}
    & 
  \accesspatternshape
  {a_0, \dots, a_n, b_0,\dots, b_m}
  {2, c_0, \dots, c_p}
    \\
    
    \texttt{windows} 
    &
    \accesspatternshape{a_0, \dots, a_m}{b_0, \dots, b_n}, \newline
    window shape $(w_0, \dots, w_n)$,
    strides $(s_0, \dots, s_n)$
    &
    \accesspatternshape{a_0, \ldots, a_m, b'_0, \dots, b'_n}{w_0, \dots, w_n},\newline
    where $b'_i = \lceil (b_i - (k_i - 1)) / s_i \rceil $\\
    
    \texttt{slice} &
    \accesspatternshape{a_0, \dots }{\dots, a_n}, \newline
    dimension index $d$, bounds $[l, h)$
    &
    \accesspatternshape{a'_0, \dots }{\dots, a'_n} \newline
    with $a'_i = a_i$ except $a'_d = h - l$
    \\
    
    \texttt{squeeze} &
    \accesspatternshape{a_0, \dots }{\dots, a_n}, index $d$ where $a_d = 1$
    &
    \accesspatternshape{a_0, \dots }{\dots, a_n} with $a_d$ removed
    \\
    
    \texttt{flatten} &
    \accesspatternshape{a_0,\dots,a_m}{b_0,\dots,b_n} &
    \accesspatternshape{a_0 \cdots a_m}{b_0 \cdots b_n} \\
    
    \texttt{reshape} &
    \accesspatternshape{a_0,\dots,a_m}{b_0,\dots,b_n},\newline
    access pattern shape literal
    \accesspatternshape{c_0,\dots,c_p}{d_0,\dots,d_q}&
    
    \accesspatternshape{c_0,\dots,c_p}{d_0,\dots,d_q},\newline
    if $a_0 \cdots a_m = c_0 \cdots c_p$
    and $b_0 \cdots b_n = d_0 \cdots d_q$\\
    
    \texttt{pair}&
    two access patterns of shape
  \accesspatternshape
  {a_0, \dots}{\dots, a_n} &
  \accesspatternshape
  {a_0, \dots}{2, \dots, a_n}
    \\
    
    \end{tabularx}
\end{table*}






This section details \g's implementation,
  focusing on its core abstraction,
  \textit{access patterns}.
We use Section~\ref{sec:matmul}'s
  matrix multiplication as a
  running example throughout.

\subsection{Access Patterns}


Access patterns encode common
  tensor IR patterns where
  some tensor dimensions
  are \textit{iterated over} (accessed)
  while others are \textit{computed on}.\footnote{
    This is similar to NumPy's concept of \textit{universal functions.}}
Section~\ref{sec:matmul}'s \tcd{matMul} example
  \textit{iterates over} dimension 0 of input $P$,
  while \textit{computing on} dimension 1,
  effectively viewing $P$ as a 1D vector of 1D vectors.

Access patterns are specified by their \textit{shape} ---
  a pair of tuples of positive integers $(S_A, S_C)$.
An access pattern of shape $(S_A, S_C)$ is, in turn, a
  tensor $T$ whose shape is given by the
  concatenation of the access pattern shape tuples
  $S_A \,\mcd{++}\, S_C$; we refer to
  $S_A$ and $S_C$ as the \textit{access} and
  \textit{compute}
  dimensions of $T$, respectively.

Access patterns represent the view of an
  $(|S_A| + |S_C|)$--dimensional tensor
  as a tensor of shape $S_A$,
  each of whose elements has shape $S_C$.
For an access pattern $T$ of shape $(S_A, S_C)$
  where $|S_A| = n_A$, we use the syntax
  \tcd{(access $T$ $n_A$)} to represent $T$ in \g.
For example, if a 2D matrix $T$ has shape $(m, n)$,
  then the \g expression \tcd{(access $T$ 1)}
  yields an access pattern of shape $((m), (n))$.

  
The matrix multiplication example
  from Section~\ref{sec:matmul}
  directly accesses the rows of $P$,
  but uses \tcd{trans2} to iterate over
  the columns of $Q$.
Instead of requiring an explicit
  transpose operator, \g provides
  access pattern \textit{transformers}.
  
\subsection{Access Pattern Transformers}

Access pattern transformers 
  manipulate one
  or more access patterns
  to produce a new access pattern,
  allowing \g
  to support more complex patterns
  like
  slicing,
  transposing,
  and interleaving.
  Table~\ref{tab:access-pattern-transformers}
  lists \g's transformers.
  
To produce an access pattern
  representing
  the columns of $Q$
  for matrix multiplication,
  we employ
  the \texttt{transpose}
  transformer.
It takes an access pattern
  and a list of dimension indices,
  and rearranges
  the dimensions 
  of the access pattern
  in the order specified by the indices.
If $Q$ has shape $(N, O)$,
  \texttt{(transpose (access $Q$ 1) (list 1 0))}
  produces
  an access pattern
  of shape
  \accesspatternshape{O}{N}.
  
The \texttt{cartProd} transformer
  takes access patterns
  of shapes
  \accesspatternshape{a_0, \dots, a_n}{c_0, \dots, c_p}
  and 
  \accesspatternshape{b_0, \dots, b_m}{c_0, \dots, c_p}
  respectively, and produces 
  an access pattern of the shape
  \accesspatternshape
    {a_0, \dots, a_n, b_0,\dots, b_m}
    {2, c_0, \dots, c_p},
  where $(2, c_0, \dots, c_p)$
  represents a 2-tuple
  of the input access patterns'
  compute dimensions.
The access dimensions
  of the input access patterns
  are simply concatenated.
In the matrix multiplication example,
  the Cartesian product
  of the rows of $P$
  with the columns of $Q$
  is an access pattern
  of shape
  \accesspatternshape{M,O}{2, N},
  where the second shape
  represents a 2-tuple
  of a row from $P$
  with a column from $Q$.

We have nearly re-implemented
  matrix multiplication example
  in \g.
The final step
  is to implement the dot product, for which
  \g uses 
  access pattern \textit{operators}.
  
\subsection{Access Pattern Operators}

\begin{table}
    \centering
    \caption{\g's access pattern operators.}
    \label{tab:operators}
    \begin{tabularx}{\linewidth}{lXX}
    Operator & Type & Description\\
    \hline
    \texttt{reduceSum} & $(\dots) \rightarrow ()$ &
    sum values
    \\
    
    \texttt{reduceMax} & $(\dots) \rightarrow ()$&
    max of all values\\
    
    \texttt{dotProd} &
    $(t,s_0, \dots, s_n)\rightarrow ()$ &
    eltwise mul; sum
    \\
    

    \end{tabularx}
\end{table}

\textit{Operators}
  are the only \g
  constructs
  which
  perform computation.
They are invoked only
  in \texttt{compute} expressions,
  which map the operator
  over the compute dimensions
  of an access pattern.
For an input access pattern
  $A$
  of shape
  \accesspatternshape
  {s_0, \dots, s_{m-1}}
  {s_m, \dots, s_{n}},
  and an operator
  $f$
  with type
  $(s_m,\dots,s_n)
  \rightarrow
  (s'_{m'}, \dots, s'_{n'})$,
  the result of
  \texttt{(compute $f$ $A$)}
  will have shape
  \accesspatternshape
  {s_0, \dots, s_{m-1}}
  {s'_{m'}, \dots, s'_{n'}};
  that is, a \tcd{compute}
  expression
  cannot change
  the access dimensions
  of the input access pattern.
Table \ref{tab:operators}
  lists
  the operators
  in \g{}.
  
Recall where we are
  in converting
  our matrix multiplication
  example:
  we have accessed the rows of $P$
  and the columns of $Q$
  and taken their Cartesian product,
  resulting in an access pattern
  of shape
  \accesspatternshape
  {M, O}{2, N},
  and we need now
  to compute the dot product
  of these row-column
  pairs.
In \g,
  the \texttt{dotProd}
  operator
  (see Table~\ref{tab:operators})
  does just that.
To compute the dot product
  over our row-column pairs,
  we need only to apply
  \texttt{compute dotProd}
  to our access pattern,
  to produce an access pattern
  with final shape
  \accesspatternshape
  {M, N}{}.
The entire \g
  specification
  of matrix multiplication
  is shown in Figure \ref{fig:mat-mat-mult}.
  
\definecolor{gray}{Gray}{5} 
  
\begin{figure*}
\begin{minipage}{.54\textwidth}
\begin{subfigure}{\textwidth}
\begin{lstlisting}[escapechar=!]
(transpose                   !\color{gray}; \hspace{2mm}\accesspatternshape{N, O, H', W'}{}!
 (squeeze                    !\color{gray}; \hspace{2mm}\accesspatternshape{N, H', W', O}{}!
  (compute dotProd           !\color{gray}; \hspace{2mm}\accesspatternshape{N, 1, H', W', O}{}!
   (cartProd                 !\color{gray}; \hspace{2mm}\accesspatternshape{N, 1, H', W', O}{2, C, K_h, K_w}!
    (windows                 !\color{gray}; \hspace{2mm}\accesspatternshape{N, 1, H', W'}{C, K_h, K_w}!
     (access activations 1)  !\color{gray}; \hspace{2mm}\accesspatternshape{N}{C,H,W}!
     (shape C Kh Kw)
     (shape 1 Sh Sw))
    (access weights 1)))     !\color{gray}; \hspace{2mm}\accesspatternshape{O}{C, K_h, K_w}!
  1)
 (list 0 3 1 2))
 
     \end{lstlisting}
       \vspace{-1.5em}
    \subcaption{2D convolution.
    }
    \label{fig:conv2d}
\end{subfigure}
\end{minipage}
\begin{minipage}{.45\textwidth}

\begin{subfigure}{\textwidth}
\begin{lstlisting}[escapechar=!]
(compute dotProd          !\color{gray}; \hspace{2mm}\accesspatternshape{M, O}{}!
 (cartProd                !\color{gray}; \hspace{2mm}\accesspatternshape{M, O}{2, N}!
  (access activations 1)  !\color{gray}; \hspace{2mm}\accesspatternshape{M}{N}!
  (transpose              !\color{gray}; \hspace{2mm}\accesspatternshape{O}{N}!
   (access weights 1)     !\color{gray}; \hspace{2mm}\accesspatternshape{N}{O}!
   (list 1 0))))
  \end{lstlisting}
  \vspace{-1.5em} 
  \subcaption{Matrix multiplication.}
  \label{fig:mat-mat-mult}
\end{subfigure}

\begin{subfigure}{\textwidth}
\begin{lstlisting}[escapechar=!]
(compute reduceMax       !\color{gray}; \accesspatternshape{N,C,H',W'}{}!
 (windows                !\color{gray}; \accesspatternshape{N,C,H',W'}{K_h, K_w}!
  (access activations 2) !\color{gray}; \accesspatternshape{N, C}{H, W}!
  (shape Kh Kw)
  (shape Sh Sw)))
\end{lstlisting}
  \vspace{-1em} 
  \subcaption{Max pooling.}
  \label{fig:maxpool-code}
\end{subfigure}

\end{minipage}
\caption{Common tensor kernels from machine learning expressed in \g. Lines containing access patterns are annotated with their access pattern shape.
$N$ is batch size; $H$/$W$ are spatial dimension sizes; $C$/$O$ are input/output channel count; $K_h$/$K_w$ are filter height/width; $S_h$/$S_w$ are strides.
}
\label{fig:all-kernels}
\end{figure*}

\section{Case Studies}
\label{sec:case-studies}

To demonstrate \g's utility,
  we first show how it enables
  concise specifications of several
  critical ML kernels
  (Section~\ref{section:representing-kernels}).
We then show how
  \g's pure, binder-free
  representation enables mapping kernels
  to an example accelerator via
  direct application of generic rewrite rules
  (Section~\ref{sec:case-study-tensorization}).
Finally,
  we highlight how \g
  enables the
  flexible mapping of
  larger, more diverse kernels
  to our accelerator,
  utilizing the power
  of equality saturation
  to automatically discover
  a variety of program transformations.
Specifically,
  we show how \g can automatically
  map convolutions to matrix multiplications
  (Section~\ref{sec:discovering-im2col})
  and automatically
  map large matrix multiplications into a
  sequence of smaller matrix multiplications
  (Section~\ref{sec:case-study-blocking}).

\subsection{Representation of Common ML Kernels}
\label{section:representing-kernels}

Figure~\ref{fig:all-kernels}
  lists the \g specifications
  of three common ML kernels:
  2D convolution,
  matrix multiplication,
  and max pooling.
Below, we discuss
  the specifications of 
  2D convolution
  and max pooling;
  see 
  Section~\ref{sec:glenside}
  for a description 
  of matrix multiplication.
  
\subsubsection*{2D Convolution}

2D convolution (\ctd{})
  is a core kernel
  in deep learning,
  defined element-by-element 
  over tensors storing
  activations $A$,
  strides $S$, and
  weights $W$ as: 
\begin{equation*}
\begin{split}
\mbox{out}&[n, o, x, y] =\\
\sum_{dx, dy, c}&
    (A[n, c, S[0] \cdot x  + dx, S[1] \cdot y + dy] \
    \cdot W[o, c, dx, dy])
\end{split}
\end{equation*}
where
  $n$ indexes the output batch,
  $o$ indexes output channels,
  $x$/$y$ index spatial dimensions,
  $dx$/$dy$ index
    the convolutional window spatial dimensions,
  and $c$ indexes input channels.
2D convolution
  slides each of the $o$
  filters
  of shape $(c, dx, dy)$
  through each possible
  $(c, dx, dy)$--shaped window
  of the input images.
At each of these locations,
  an elementwise multiplication
  and reduction sum
  is computed.

The \g specification
  of \ctd{}
  is shown in 
  Figure \ref{fig:conv2d}.
We access
  the \texttt{weights}
  as a vector of $O$ filters
  and the \texttt{activations}
  as a vector of $N$ images.
We leave the filters as they are,
  but form windows
  of shape
  $(C, K_h, K_w)$
  over the activations
  using the \texttt{windows}
  access pattern transformer
  (Table~\ref{tab:access-pattern-transformers}).
This produces an access pattern
  of shape
  \accesspatternshape
  {N, 1, H', W'}
  {C, K_h, K_w},
  i.e.,
  a batch of ``images''
  of new spatial shape
  $(H', W')$,
  where every location
  is a window of
  the original input.
Finally,
  we take the Cartesian product
  of the filters
  and the windows,
  compute their dot product,
  and \texttt{squeeze} and \texttt{transpose}
  the output
  into the correct layout.

\subsubsection*{Max Pooling}

Max pooling, commonly used in ML
  to condense intermediate activations,
  is defined as:
\begin{equation*}
\begin{split} 
\mbox{out}&[n, c, x, y] =\\
\max_{dx, dy}&
           (\mbox{activations}[n, c,
                       \mbox{strides}[0] \cdot x  + dx,  
                       \mbox{strides}[1] \cdot y + dy])
\end{split}
\end{equation*}

Max pooling 
  slides a window
  of shape $(dx, dy)$
  over all possible locations
  within the spatial (i.e.,~$x$ and $y$)
  dimensions.
At each window location,
  it reduces the window
  to a scalar
  with the $\max$ operator.
The \g specification merely applies
  \texttt{reduceMax} 
  over each two-dimensional window.
  
  
\subsubsection*{Discussion}\label{section:kernel-implementation-discussion}

\g separates
  the \textit{computation}
  from the \textit{data access patterns}
  in these kernels while exposing
  the simplicity of their computation---%
  and the relative complexity
  of their data access.
In all three kernels,
  the computation can be described
  with a single operator;
  most of the specification
  entails
  setting up the data access pattern.

Furthermore,
  \g exposes similar structure
  between kernels;
  for example,
  both \ctd
  and matrix multiplication
  feature the expression
  \tcd{(compute dotProd (cartProd ...))}.
At their core, these kernels
  are performing the same computation,
  but with different patterns
  of data access.
In Section~\ref{sec:discovering-im2col},
  we exploit this similarity in structure
  when mapping kernels to hardware.
  
These kernels highlight the expressive power
  of access patterns.
Consider the use of 
  \tcd{windows}
  in \ctd
   and max pooling.
Both kernels
  form windows
  differently:
  \ctd forms three-dimensional
  windows
  over the channels, height, and width
  dimensions,
  while max pooling forms two-dimensional windows
  over the height and width.
Rather than passing configuration parameters to \tcd{windows},
  \g attaches this information to the tensors themselves.
  
  
\begin{figure}
\begin{lstlisting}[escapechar=!]
(compute dotProd (cartProd ?a0 ?a1)) !$\Longrightarrow$!
  (systolicArray ?rows ?cols
    ?a0 (access (transpose ?a1 (list 1 0)) 0))
!$\textrm{where \texttt{?a0} is of shape $((\mbox{\texttt{?batch}}), (\mbox{\texttt{?rows}}))$}$!
  !$\textrm{and \texttt{?a1} is of shape $((\mbox{\texttt{?cols}}), (\mbox{\texttt{?rows}}))$}$!
\end{lstlisting}
\justify
\caption{Our rewrite rewriting matrix multiplication to a systolic array invocation.}
    \label{fig:systolic-array-rewrite}
\end{figure}
  
\begin{figure*}
\begin{lstlisting}[escapechar=!]
                                                    ?a !$\Longrightarrow$! (reshape (flatten ?a) ?shape) 
(cartProd (reshape ?a0 ?shape0) (reshape ?a1 ?shape1)) !$\Longrightarrow$! (reshape (cartProd ?a0 ?a1) ?newShape)
                 (compute dotProd (reshape ?a ?shape)) !$\Longrightarrow$! (reshape (compute dotProd ?a) ?newShape)
\end{lstlisting}
\caption{Rewrites enabling the discovery of the \itc transformation.}
\label{fig:im2col-rewrites}
\end{figure*}

\subsection{Mapping \tcd{matMul} to Accelerators}
\label{sec:case-study-tensorization}

\g can be used to uncover opportunities
  to invoke accelerator components.
Consider a 
  weight-stationary systolic array,
  a common matrix multiplication
  architecture.
A weight-stationary
  systolic array
  with $r$ rows
  and $c$ columns
  takes two lists
  of length-$r$ vectors
  (the activations
    and weights, respectively),
  pairing each vector
  from one list
  with each vector
  from the other,
  and computes a dot product
  over each pair.
The second list
  contains $c$ vectors,
  while the first
  can be of any length.
  
\g's purity
  allows us to implement this hardware mapping task
  using a term rewriting system,
  in which we rewrite a matching program pattern
  to an invocation of our systolic array.
Our rewrite is shown in 
  Figure~\ref{fig:systolic-array-rewrite},
  mimicking
  \tcd{egg}'s rewrite syntax.
Tokens starting with a question mark
  (such as \texttt{?a0} in 
  Figure~\ref{fig:systolic-array-rewrite})
  are variables in the pattern,
  bound by the left-hand side (LHS),
  and then used on the right-hand side (RHS).
\tcd{egg} also allows for
  conditions on rewrites,
  which we print below our rewrites.

To design our rewrite,
  we first must design
  the LHS
  to match program patterns
  that resemble the data access pattern
  and compute pattern
  of our systolic array.
\g is eminently suitable for this task,
  as it can express
  exactly the data access
  and computation pattern
  we described
  for the systolic array.
Pairing all vectors from one list
  with all vectors from another
  and computing the dot product
  of the pairs
  is represented as
  \tcd{(compute dotProd (cartProd ?a0 ?a1))},
  binding
  \tcd{?a0}
  and \tcd{?a1}
  to the input access patterns.
We encode
  the systolic array's
  constraints
  on the input shapes
  as a condition on the rewrite.
Patterns which match the LHS
  are mapped to the RHS;
  in this case, we introduce a new
  \tcd{systolicArray} construct
  to represent the functioning of our systolic array.
The shape of the systolic array 
  is given by the \tcd{?rows} and \tcd{?cols}
  parameters,
  and the inputs are given
  as access patterns.
Note how we also transform
  the second access pattern
  to more accurately convey
  how the actual systolic array
  hardware
  accesses the weight tensor:
  it reads it all at once
  (hence, \tcd{(access ... 0)}),
  and expects it to be laid out
  in transposed form
  in memory.
This added information---%
  enabled by \g's access patterns---%
  provides richer data layout information,
  potentially helping future rewrites
  or code generation steps.
 
\subsection{Flexible Mapping: Discovering \itc{}}\label{sec:discovering-im2col}

%

\begin{figure}
\begin{lstlisting}[escapechar=!]
(transpose                   
 (squeeze                    
  (reshape           !\color{gray}; \hspace{2mm}\accesspatternshape{N,1,H',W',O}{}!
   (compute dotProd  !\color{gray}; \hspace{2mm}\accesspatternshape{N \cdot 1 \cdot H' \cdot W',O}{}!          
    (cartProd                 
     (flatten        !\color{gray}; \hspace{2mm}\accesspatternshape{N \cdot 1 \cdot H' \cdot W'}{C \cdot K_h \cdot K_w}!
      (windows (access activations 1)  
               (shape C Kh Kw) (shape 1 Sh Sw)))
     (flatten        !\color{gray}; \hspace{2mm}\accesspatternshape{O}{C \cdot K_h \cdot K_w}!
      (access weights 1))))
   ?shape) 1) (list 0 3 1 2))
     \end{lstlisting}
     \vspace{-1em}
    \caption{An \tcd{im2col}-transformed 
      \ctd,
    after the application of the rewrites
    in Figure~\ref{fig:im2col-rewrites}
    and just before the application
    of the systolic array rewrite.}
    \label{fig:conv2d-im2col-rewritten}
\end{figure}
  
The \tcd{im2col} transformation
  is a data layout optimization
  which enables computing \ctd
  on matrix multiplication hardware.
The transformation
  involves instantiating
  the convolutional windows
  over the input activations
  directly in memory~\cite{im2col}.
This leads to data duplication,
  but the resulting speedup
  more than offsets that overhead.
In this case study,
  we show how a few
  general rewrites
  within \g
  lead to the 
  \textit{automatic rederivation}
  of the 
  \tcd{im2col} transformation.

\g's representation underscores
  the structural similarity
  between \ctd and matrix multiplication,
  reflected also by the shared
  \tcd{(compute dotProd (cartProd ...))}
  between \ctd 
  and the LHS of the systolic array rewrite
  in Figure~\ref{fig:systolic-array-rewrite}.
Using this rewrite 
  on \ctd would permit mapping it to the systolic array; 
  however, 
  the restrictions
  on the shape of
  \texttt{?a0}
  and \texttt{?a1}
  prevent its application.
The systolic array has 
  an activation access pattern
  of shape \accesspatternshape
  {a}{b}
  and a weight access pattern
  of shape \accesspatternshape
  {c}{d},
  while \ctd operates over
  access patterns
  of shape \accesspatternshape
  {N, 1,H',W'}{C, K_h, K_w}
  and
  of \accesspatternshape
  {O}{C, K_h, K_w},
  respectively.
Transforming the access pattern
  into a lower-dimensional form
  would enable the systolic array rewrite.
  
Figure~\ref{fig:im2col-rewrites}
  shows the rewrites
  which enable this transformation.
We call the first rewrite
  an 
  \textit{exploratory} rewrite
  as it
  optimistically matches
  any access pattern expression.
It flattens 
  an access pattern
  and immediately reshapes it
  back to its original shape, thus preserving equality
  (see Table~\ref{tab:access-pattern-transformers}
  for formal definitions).
This exploratory rewrite introduces the flattening
  necessary
  to convert the higher-dimensional access patterns
  of \ctd
  into the access patterns
  matched by the systolic array rewrite.
However, the \tcd{reshape} operator
  will still need to be moved 
  before we can fire 
  Figure~\ref{fig:systolic-array-rewrite}'s
  systolic array rewrite.
The second and third rewrites
  in Figure~\ref{fig:im2col-rewrites}
  take care of this;
  they implement \textit{composition commutativity}
  of \tcd{reshape}
  with \tcd{cartProd} and \tcd{compute dotProd},
  which ``bubble'' \tcd{reshape} operators
  up and out of expressions.
These rewrites
  express general properties of these operators
  and are not specific
  to this task.
  
These three rewrites work in concert 
  to map \ctd
  to a systolic array.
First,\footnote{
Since equality saturation 
  explores rewrites non-destructively, 
  the rewriting order here
  is purely for explanatory purposes.
}
  the exploratory rewrite
  flattens and reshapes
  all access pattern expressions.
This includes the inputs
  to \ctd's \tcd{cartProd}
  subexpression,
  which are flattened
  to shapes
  \accesspatternshape
  {N \cdot 1 \cdot H' \cdot W'}{C \cdot K_h \cdot K_w}
  and
  \accesspatternshape
  {O}{C \cdot K_h \cdot K_w}
  and reshaped
  back to their original shapes.
Next,
  the composition commutativity rewrites
  for \tcd{cartProd} 
  and
  \tcd{compute dotProd}
  fire in sequence,
  bubbling the \tcd{reshape} up
  through the 
  \tcd{cartProd} 
  and \tcd{dotProd} 
  expressions (shown in Figure~\ref{fig:conv2d-im2col-rewritten}).
Finally,
  the systolic array rewrite
  completes the \tcd{im2col} transform.
\g's equality saturation based rewrite engine
  discovers these rewrites
  because the exploratory rewrite 
  fires on every term
  and no rewrites are missed
  due to the absence of phase ordering.

This example highlights how,
  \textit{with straightforward,
  generally applicable rewrites
  defined over \g},
  equality saturation
  can emergently discover useful transformations
  that previously required
  expert insights to apply.

\subsection{Flexible Mapping: \tcd{matMul} Blocking}
\label{sec:case-study-blocking}


%

\begin{figure}
\begin{lstlisting}[escapechar=!]
?a !$\Longrightarrow$! (concat (slice ?a ?dim ?b0 ?b1)
               (slice ?a ?dim ?b1 ?b2) ?dim)
               
(cartProd ?a (concat ?b0 ?b1 ?dim)) !$\Longrightarrow$!
 (concat (cartProd ?a ?b0) (cartProd ?a ?b1) ?newDim)
!$\textrm{if \texttt{?dim} is an access dimension}$!
 
(cartProd (concat ?a0 ?a1 ?dim0) 
          (concat ?a2 ?a3 ?dim1)) !$\Longrightarrow$!
 (concat (cartProd ?a0 ?a2) 
         (cartProd ?a1 ?a3) ?newDim)
!$\textrm{if \texttt{?dim0} and \texttt{?dim1} are the same shape dimension}$!

(compute dotProd (concat ?a0 ?a1 ?dim)) !$\Longrightarrow$!
 (concat (compute dotProd ?a0)
         (compute dotProd ?a1) ?dim)
!$\textrm{if \texttt{?dim} is an access dimension}$!

(compute dotProd (concat ?a0 ?a1 ?dim)) !$\Longrightarrow$!
 (compute reduceSum (pair (compute dotProd ?a0)
                          (compute dotProd ?a1)))
!$\textrm{if \texttt{?dim} is a shape dimension}$!
\end{lstlisting}
\justify
\vspace{-1em}
\caption{
Rewrites for blocking \tcd{matMul}.
}
\label{fig:all-blocking-rewrites}
\end{figure}

\begin{figure}
\begin{lstlisting}[escapechar=!]
(concat                          !\color{gray}; \hspace{2mm}\accesspatternshape{32,32}{}!
 (concat                         !\color{gray}; \hspace{2mm}\accesspatternshape{16,32}{}!
  (compute reduceSum             !\color{gray}; \hspace{2mm}\accesspatternshape{16,16}{}!
   (pair                         !\color{gray}; \hspace{2mm}\accesspatternshape{16,16}{2}!
    (compute dotProd             !\color{gray}; \hspace{2mm}\accesspatternshape{16,16}{}!
     (cartProd                   !\color{gray}; \hspace{2mm}\accesspatternshape{16,16}{2, 16}!
      (slice                     !\color{gray}; \hspace{2mm}\accesspatternshape{16}{16}!
       (slice (access activations 1) 0 0 16) 1 0 16)
      (transpose                 !\color{gray}; \hspace{2mm}\accesspatternshape{16}{16}!
       (slice 
        (slice (access weights 1) 0 0 16) 1 0 16)
       (list 1 0))))
    (compute dotProd             !\color{gray}; \hspace{2mm}\accesspatternshape{16,16}{}!
     (cartProd                   !\color{gray}; \hspace{2mm}\accesspatternshape{16,16}{2, 16}!
      (slice                     !\color{gray}; \hspace{2mm}\accesspatternshape{16}{16}!
       (slice (access activations 1) 0 16 32) 1 0 16)
      (transpose                 !\color{gray}; \hspace{2mm}\accesspatternshape{16}{16}!
       (slice                    !\color{gray}; \hspace{2mm}\accesspatternshape{16}{16}!
        (slice (access weights 1) 0 16 32) 1 0 16)
       (list 1 0)))))) ...
  \end{lstlisting}
  \vspace{-2em}
  \caption{A $32\times32$ \texttt{matMul}
  blocked into $16\times16$ \texttt{matMul}s.
  Only two of the eight total multiplications are shown.
  }
  \label{fig:matmul-rewritten}
  \vspace{-1em}
\end{figure}
Equality saturation 
  can also be used with \g 
  to emergently discover a
  matrix multiplication
  blocking scheme.
Matrix multiplication blocking
  is the common strategy
  of breaking up a single, large
  matrix multiplication
  into smaller multiplications,
  by multiplying subsets
  of the input matrices
  and assembling the results
  to form the output matrix.
This is essential in practice,
  as systolic arrays are small
  (often between $16\times16$ and $256\times256$)
  while matrices in ML and HPC applications
  can be much larger.

As in Section~\ref{sec:discovering-im2col},
  this transformation follows
  from an exploratory rewrite
  and associated ``cleanup'' rewrites.
The exploratory rewrite used for blocking
  is shown at the top of Figure~\ref{fig:all-blocking-rewrites}.
Given an access pattern,
  this rewrite slices the access pattern
  into two pieces
  along a dimension
  and then concatenates them back together.
The dimension
  as well as the division strategy
  are configurable.
For this example,
  we assume for simplicity
  that we run this rewrite
  on every available dimension,
  that we divide each dimension
  perfectly in half,
  and that all dimensions are powers of 2 in size.
Figure~\ref{fig:all-blocking-rewrites} gives rewrites
  for bubbling the introduced
  \texttt{concat}
  operators up through the expression,
  namely the compositional commutativity
  of \tcd{concat}
  with \tcd{cartProd}
  and \tcd{compute dotProd}.
Starting from the matrix multiplication
  in Figure \ref{fig:mat-mat-mult},
  assuming input shapes of $(32,32)$,
  the exploratory rewrite first slices and concatenates
  the access patterns
  at the input of \tcd{cartProd}.
Then, using the commutativity rewrites,
  the resulting \tcd{concat}s
  are bubbled up
  to produce the final expression
  in Figure~\ref{fig:matmul-rewritten}.
The effect of these rewrites
  is that the single 
  $32\times 32$ \tcd{matMul}
  becomes eight separate $16\times 16$ \tcd{matMul}s,
  which are
  summed
  and concatenated
  to form the full output matrix.
This case study demonstrates
  yet again
  that \g's expressiveness
  allows a small set
  of rewrites
  to produce interesting and useful
  emergent transformations.
  

\section{Conclusion}
\label{sec:conclusion}

In this paper,
  we proposed \textit{access patterns} as an
  abstraction to enable equality saturation
  style term rewriting for low-level
  tensor program mapping to hardware accelerators.
Crucially, access patterns support
  specifying and composing
  higher-order, arbitrary dimension
  tensor operators without the need
  for binding structures like anonymous functions
  or index notation.
We demonstrated the potential utility of
  access patterns in the \g IR through
  case studies showing how rewrites in \g
  can automatically uncover common
  layout transformations like \tcd{im2col}
  used for accelerator mapping.
We are excited for the community
  to join in further exploring the potential
  applications of access patterns and to
  build additional optimizations on
  \g's foundations.

\begin{acks}
This work
  was sponsored by
  the \grantsponsor{}{Real Time Machine Learning (RTML)}{https://www.darpa.mil/program/real-time-machine-learning}
  DARPA project,
  and the
\grantsponsor{}{Applications Driving Architectures (ADA)}{https://adacenter.org/}
and
\grantsponsor{}{Center for Research in Intelligent Storage and Processing in Memory (CRISP)}{https://crisp.engineering.virginia.edu/}
JUMP  Research Centers,
co-sponsored by SRC and DARPA.
We thank Chandrakana Nandi
  for her extensive and enthusiastic
  editing.
We also thank 
  Sorawee Porncharoenwase,
  Luis Vega, and
  Max Willsey
  for their helpful comments.
\end{acks}

\bibliographystyle{ACM-Reference-Format}
\bibliography{main}


\begin{thebibliography}{46}


\ifx \showCODEN    \undefined \def \showCODEN     #1{\unskip}     \fi
\ifx \showDOI      \undefined \def \showDOI       #1{#1}\fi
\ifx \showISBNx    \undefined \def \showISBNx     #1{\unskip}     \fi
\ifx \showISBNxiii \undefined \def \showISBNxiii  #1{\unskip}     \fi
\ifx \showISSN     \undefined \def \showISSN      #1{\unskip}     \fi
\ifx \showLCCN     \undefined \def \showLCCN      #1{\unskip}     \fi
\ifx \shownote     \undefined \def \shownote      #1{#1}          \fi
\ifx \showarticletitle \undefined \def \showarticletitle #1{#1}   \fi
\ifx \showURL      \undefined \def \showURL       {\relax}        \fi
\providecommand\bibfield[2]{#2}
\providecommand\bibinfo[2]{#2}
\providecommand\natexlab[1]{#1}
\providecommand\showeprint[2][]{arXiv:#2}

\bibitem[\protect\citeauthoryear{Abadi, Barham, Chen, Chen, Davis, Dean, Devin,
  Ghemawat, Irving, Isard, Kudlur, Levenberg, Monga, Moore, Murray, Steiner,
  Tucker, Vasudevan, Warden, Wicke, Yu, and Zheng}{Abadi et~al\mbox{.}}{2016}]%
        {tensorflow}
\bibfield{author}{\bibinfo{person}{Martin Abadi}, \bibinfo{person}{Paul
  Barham}, \bibinfo{person}{Jianmin Chen}, \bibinfo{person}{Zhifeng Chen},
  \bibinfo{person}{Andy Davis}, \bibinfo{person}{Jeffrey Dean},
  \bibinfo{person}{Matthieu Devin}, \bibinfo{person}{Sanjay Ghemawat},
  \bibinfo{person}{Geoffrey Irving}, \bibinfo{person}{Michael Isard},
  \bibinfo{person}{Manjunath Kudlur}, \bibinfo{person}{Josh Levenberg},
  \bibinfo{person}{Rajat Monga}, \bibinfo{person}{Sherry Moore},
  \bibinfo{person}{Derek~G. Murray}, \bibinfo{person}{Benoit Steiner},
  \bibinfo{person}{Paul Tucker}, \bibinfo{person}{Vijay Vasudevan},
  \bibinfo{person}{Pete Warden}, \bibinfo{person}{Martin Wicke},
  \bibinfo{person}{Yuan Yu}, {and} \bibinfo{person}{Xiaoqiang Zheng}.}
  \bibinfo{year}{2016}\natexlab{}.
\newblock \showarticletitle{TensorFlow: A system for large-scale machine
  learning}. In \bibinfo{booktitle}{\emph{12th USENIX Symposium on Operating
  Systems Design and Implementation (OSDI 16)}}. \bibinfo{pages}{265--283}.
\newblock
\urldef\tempurl%
\url{https://www.usenix.org/system/files/conference/osdi16/osdi16-abadi.pdf}
\showURL{%
\tempurl}


\bibitem[\protect\citeauthoryear{Anderson, Adams, Ma, Li, and
  Ragan-Kelley}{Anderson et~al\mbox{.}}{2020}]%
        {anderson2020learning}
\bibfield{author}{\bibinfo{person}{Luke Anderson}, \bibinfo{person}{Andrew
  Adams}, \bibinfo{person}{Karima Ma}, \bibinfo{person}{Tzu-Mao Li}, {and}
  \bibinfo{person}{Jonathan Ragan-Kelley}.} \bibinfo{year}{2020}\natexlab{}.
\newblock \bibinfo{title}{Learning to Schedule Halide Pipelines for the GPU}.
\newblock
\newblock
\showeprint[arxiv]{2012.07145}~[cs.PL]


\bibitem[\protect\citeauthoryear{Baader and Nipkow}{Baader and Nipkow}{1998}]%
        {baader1998term}
\bibfield{author}{\bibinfo{person}{Franz Baader} {and} \bibinfo{person}{Tobias
  Nipkow}.} \bibinfo{year}{1998}\natexlab{}.
\newblock \bibinfo{booktitle}{\emph{Term Rewriting and All That}}.
\newblock \bibinfo{publisher}{Cambridge University Press}.
\newblock
\urldef\tempurl%
\url{https://doi.org/10.1017/CBO9781139172752}
\showDOI{\tempurl}


\bibitem[\protect\citeauthoryear{Baghdadi, Ray, Romdhane, Sozzo, Akkas, Zhang,
  Suriana, Kamil, and Amarasinghe}{Baghdadi et~al\mbox{.}}{2019}]%
        {tiramisu}
\bibfield{author}{\bibinfo{person}{Riyadh Baghdadi}, \bibinfo{person}{Jessica
  Ray}, \bibinfo{person}{Malek~Ben Romdhane}, \bibinfo{person}{Emanuele~Del
  Sozzo}, \bibinfo{person}{Abdurrahman Akkas}, \bibinfo{person}{Yunming Zhang},
  \bibinfo{person}{Patricia Suriana}, \bibinfo{person}{Shoaib Kamil}, {and}
  \bibinfo{person}{Saman Amarasinghe}.} \bibinfo{year}{2019}\natexlab{}.
\newblock \showarticletitle{Tiramisu: A Polyhedral Compiler for Expressing Fast
  and Portable Code}.
\newblock \bibinfo{journal}{\emph{2019 IEEE/ACM International Symposium on Code
  Generation and Optimization (CGO)}} (\bibinfo{date}{Feb}
  \bibinfo{year}{2019}).
\newblock
\showISBNx{9781728114361}
\urldef\tempurl%
\url{https://doi.org/10.1109/cgo.2019.8661197}
\showDOI{\tempurl}


\bibitem[\protect\citeauthoryear{Chakravarty, Keller, Lee, McDonell, and
  Grover}{Chakravarty et~al\mbox{.}}{2011}]%
        {accelerate}
\bibfield{author}{\bibinfo{person}{Manuel M~T Chakravarty},
  \bibinfo{person}{Gabriele Keller}, \bibinfo{person}{Sean Lee},
  \bibinfo{person}{Trevor~L. McDonell}, {and} \bibinfo{person}{Vinod Grover}.}
  \bibinfo{year}{2011}\natexlab{}.
\newblock \showarticletitle{{Accelerating Haskell array codes with multicore
  GPUs}}. In \bibinfo{booktitle}{\emph{DAMP '11: The 6th workshop on
  Declarative Aspects of Multicore Programming}}. \bibinfo{publisher}{ACM}.
\newblock


\bibitem[\protect\citeauthoryear{Charlier, Feydy, Glaunès, Collin, and
  Durif}{Charlier et~al\mbox{.}}{2021}]%
        {keops}
\bibfield{author}{\bibinfo{person}{Benjamin Charlier}, \bibinfo{person}{Jean
  Feydy}, \bibinfo{person}{Joan~Alexis Glaunès},
  \bibinfo{person}{François-David Collin}, {and} \bibinfo{person}{Ghislain
  Durif}.} \bibinfo{year}{2021}\natexlab{}.
\newblock \showarticletitle{Kernel Operations on the GPU, with Autodiff,
  without Memory Overflows}.
\newblock \bibinfo{journal}{\emph{Journal of Machine Learning Research}}
  \bibinfo{volume}{22}, \bibinfo{number}{74} (\bibinfo{year}{2021}),
  \bibinfo{pages}{1--6}.
\newblock
\urldef\tempurl%
\url{http://jmlr.org/papers/v22/20-275.html}
\showURL{%
\tempurl}


\bibitem[\protect\citeauthoryear{Chellapilla, Puri, and Simard}{Chellapilla
  et~al\mbox{.}}{2006}]%
        {im2col}
\bibfield{author}{\bibinfo{person}{Kumar Chellapilla}, \bibinfo{person}{Sidd
  Puri}, {and} \bibinfo{person}{Patrice Simard}.}
  \bibinfo{year}{2006}\natexlab{}.
\newblock \showarticletitle{{High Performance Convolutional Neural Networks for
  Document Processing}}. In \bibinfo{booktitle}{\emph{{Tenth International
  Workshop on Frontiers in Handwriting Recognition}}},
  \bibfield{editor}{\bibinfo{person}{Guy Lorette}} (Ed.). {Universit{\'e} de
  Rennes 1}, \bibinfo{publisher}{{Suvisoft}}, \bibinfo{address}{La Baule
  (France)}.
\newblock
\urldef\tempurl%
\url{https://hal.inria.fr/inria-00112631}
\showURL{%
\tempurl}
\newblock
\shownote{http://www.suvisoft.com.}


\bibitem[\protect\citeauthoryear{Chen, Moreau, Jiang, Zheng, Yan, Shen, Cowan,
  Wang, Hu, Ceze, et~al\mbox{.}}{Chen et~al\mbox{.}}{2018a}]%
        {chen2018tvm}
\bibfield{author}{\bibinfo{person}{Tianqi Chen}, \bibinfo{person}{Thierry
  Moreau}, \bibinfo{person}{Ziheng Jiang}, \bibinfo{person}{Lianmin Zheng},
  \bibinfo{person}{Eddie Yan}, \bibinfo{person}{Haichen Shen},
  \bibinfo{person}{Meghan Cowan}, \bibinfo{person}{Leyuan Wang},
  \bibinfo{person}{Yuwei Hu}, \bibinfo{person}{Luis Ceze}, {et~al\mbox{.}}}
  \bibinfo{year}{2018}\natexlab{a}.
\newblock \showarticletitle{$\{$TVM$\}$: An automated end-to-end optimizing
  compiler for deep learning}. In \bibinfo{booktitle}{\emph{13th $\{$USENIX$\}$
  Symposium on Operating Systems Design and Implementation ($\{$OSDI$\}$ 18)}}.
  \bibinfo{pages}{578--594}.
\newblock


\bibitem[\protect\citeauthoryear{Chen, Moreau, Jiang, Zheng, Yan, Shen, Cowan,
  Wang, Hu, Ceze, Guestrin, and Krishnamurthy}{Chen et~al\mbox{.}}{2018b}]%
        {tvm}
\bibfield{author}{\bibinfo{person}{Tianqi Chen}, \bibinfo{person}{Thierry
  Moreau}, \bibinfo{person}{Ziheng Jiang}, \bibinfo{person}{Lianmin Zheng},
  \bibinfo{person}{Eddie Yan}, \bibinfo{person}{Haichen Shen},
  \bibinfo{person}{Meghan Cowan}, \bibinfo{person}{Leyuan Wang},
  \bibinfo{person}{Yuwei Hu}, \bibinfo{person}{Luis Ceze},
  \bibinfo{person}{Carlos Guestrin}, {and} \bibinfo{person}{Arvind
  Krishnamurthy}.} \bibinfo{year}{2018}\natexlab{b}.
\newblock \showarticletitle{{TVM}: An Automated End-to-End Optimizing Compiler
  for Deep Learning}. In \bibinfo{booktitle}{\emph{13th {USENIX} Symposium on
  Operating Systems Design and Implementation ({OSDI} 18)}}.
  \bibinfo{publisher}{{USENIX} Association}, \bibinfo{address}{Carlsbad, CA},
  \bibinfo{pages}{578--594}.
\newblock
\showISBNx{978-1-931971-47-8}
\urldef\tempurl%
\url{https://www.usenix.org/conference/osdi18/presentation/chen}
\showURL{%
\tempurl}


\bibitem[\protect\citeauthoryear{Chen, Zheng, Yan, Jiang, Moreau, Ceze,
  Guestrin, and Krishnamurthy}{Chen et~al\mbox{.}}{2018c}]%
        {chen2018autotvm}
\bibfield{author}{\bibinfo{person}{Tianqi Chen}, \bibinfo{person}{Lianmin
  Zheng}, \bibinfo{person}{Eddie Yan}, \bibinfo{person}{Ziheng Jiang},
  \bibinfo{person}{Thierry Moreau}, \bibinfo{person}{Luis Ceze},
  \bibinfo{person}{Carlos Guestrin}, {and} \bibinfo{person}{Arvind
  Krishnamurthy}.} \bibinfo{year}{2018}\natexlab{c}.
\newblock \showarticletitle{Learning to Optimize Tensor Programs}. In
  \bibinfo{booktitle}{\emph{Proceedings of the 32nd International Conference on
  Neural Information Processing Systems}} (Carlsbad, CA, USA)
  \emph{(\bibinfo{series}{NIPS'18})}. \bibinfo{publisher}{USENIX Association},
  \bibinfo{address}{USA}, \bibinfo{pages}{3393–3404}.
\newblock


\bibitem[\protect\citeauthoryear{Chen and Yu}{Chen and Yu}{2020}]%
        {byoc}
\bibfield{author}{\bibinfo{person}{Zhi Chen} {and} \bibinfo{person}{Cody Yu}.}
  \bibinfo{year}{2020}\natexlab{}.
\newblock \bibinfo{title}{How to Bring Your Own Codegen to TVM}.
\newblock
  \bibinfo{howpublished}{\url{https://tvm.apache.org/2020/07/15/how-to-bring-your-own-codegen-to-tvm}}.
\newblock


\bibitem[\protect\citeauthoryear{{Chen, Yu-Hsin and Krishna, Tushar and Emer,
  Joel and Sze, Vivienne}}{{Chen, Yu-Hsin and Krishna, Tushar and Emer, Joel
  and Sze, Vivienne}}{2016}]%
        {chen2016eyeriss}
\bibfield{author}{\bibinfo{person}{{Chen, Yu-Hsin and Krishna, Tushar and Emer,
  Joel and Sze, Vivienne}}.} \bibinfo{year}{{2016}}\natexlab{}.
\newblock \showarticletitle{{Eyeriss: An Energy-Efficient Reconfigurable
  Accelerator for Deep Convolutional Neural Networks}}. In
  \bibinfo{booktitle}{\emph{{IEEE International Solid-State Circuits
  Conference, ISSCC 2016, Digest of Technical Papers}}}.
  \bibinfo{pages}{{262--263}}.
\newblock


\bibitem[\protect\citeauthoryear{Garavel, Tabikh, and Arrada}{Garavel
  et~al\mbox{.}}{2018}]%
        {garavel2018rewrite-context}
\bibfield{author}{\bibinfo{person}{Hubert Garavel},
  \bibinfo{person}{Mohammad-Ali Tabikh}, {and} \bibinfo{person}{Imad-Seddik
  Arrada}.} \bibinfo{year}{2018}\natexlab{}.
\newblock \showarticletitle{{Benchmarking Implementations of Term Rewriting and
  Pattern Matching in Algebraic, Functional, and Object-Oriented Languages -
  The 4th Rewrite Engines Competition}}. In
  \bibinfo{booktitle}{\emph{{Proceedings of the 12th International Workshop on
  Rewriting Logic and its Applications (WRLA'18)}}}.
  \bibinfo{address}{Thessaloniki, Greece}.
\newblock
\urldef\tempurl%
\url{https://hal.inria.fr/hal-01883212}
\showURL{%
\tempurl}


\bibitem[\protect\citeauthoryear{Hagedorn, Elliott, Barthels, Bodik, and
  Grover}{Hagedorn et~al\mbox{.}}{2020a}]%
        {hagedorn2020fireiron}
\bibfield{author}{\bibinfo{person}{Bastian Hagedorn},
  \bibinfo{person}{Archibald~Samuel Elliott}, \bibinfo{person}{Henrik
  Barthels}, \bibinfo{person}{Rastislav Bodik}, {and} \bibinfo{person}{Vinod
  Grover}.} \bibinfo{year}{2020}\natexlab{a}.
\newblock \bibinfo{title}{Fireiron: A Scheduling Language for High-Performance
  Linear Algebra on GPUs}.
\newblock
\newblock
\showeprint[arxiv]{2003.06324}~[cs.PL]


\bibitem[\protect\citeauthoryear{Hagedorn, Lenfers, Kundefinedhler, Qin,
  Gorlatch, and Steuwer}{Hagedorn et~al\mbox{.}}{2020b}]%
        {hagedorn2020func-high-perf}
\bibfield{author}{\bibinfo{person}{Bastian Hagedorn}, \bibinfo{person}{Johannes
  Lenfers}, \bibinfo{person}{Thomas Kundefinedhler}, \bibinfo{person}{Xueying
  Qin}, \bibinfo{person}{Sergei Gorlatch}, {and} \bibinfo{person}{Michel
  Steuwer}.} \bibinfo{year}{2020}\natexlab{b}.
\newblock \showarticletitle{Achieving High-Performance the Functional Way: A
  Functional Pearl on Expressing High-Performance Optimizations as Rewrite
  Strategies}.
\newblock \bibinfo{journal}{\emph{Proc. ACM Program. Lang.}}
  \bibinfo{volume}{4}, \bibinfo{number}{ICFP}, Article \bibinfo{articleno}{92}
  (\bibinfo{date}{Aug.} \bibinfo{year}{2020}), \bibinfo{numpages}{29}~pages.
\newblock
\urldef\tempurl%
\url{https://doi.org/10.1145/3408974}
\showDOI{\tempurl}


\bibitem[\protect\citeauthoryear{Jia}{Jia}{2014}]%
        {jia2014semantic}
\bibfield{author}{\bibinfo{person}{Yangqing Jia}.}
  \bibinfo{year}{2014}\natexlab{}.
\newblock \emph{\bibinfo{title}{Learning Semantic Image Representations at a
  Large Scale}}.
\newblock \bibinfo{thesistype}{Ph.D. Dissertation}. \bibinfo{school}{EECS
  Department, University of California, Berkeley}.
\newblock
\urldef\tempurl%
\url{http://www2.eecs.berkeley.edu/Pubs/TechRpts/2014/EECS-2014-93.html}
\showURL{%
\tempurl}


\bibitem[\protect\citeauthoryear{Jouppi, Young, Patil, Patterson, Agrawal,
  Bajwa, Bates, Bhatia, Boden, Borchers, Boyle, Cantin, Chao, Clark, Coriell,
  Daley, Dau, Dean, Gelb, Ghaemmaghami, Gottipati, Gulland, Hagmann, Ho,
  Hogberg, Hu, Hundt, Hurt, Ibarz, Jaffey, Jaworski, Kaplan, Khaitan,
  Killebrew, Koch, Kumar, Lacy, Laudon, Law, Le, Leary, Liu, Lucke, Lundin,
  MacKean, Maggiore, Mahony, Miller, Nagarajan, Narayanaswami, Ni, Nix, Norrie,
  Omernick, Penukonda, Phelps, Ross, Ross, Salek, Samadiani, Severn, Sizikov,
  Snelham, Souter, Steinberg, Swing, Tan, Thorson, Tian, Toma, Tuttle,
  Vasudevan, Walter, Wang, Wilcox, and Yoon}{Jouppi et~al\mbox{.}}{2017}]%
        {jouppi2017tpu}
\bibfield{author}{\bibinfo{person}{Norman~P. Jouppi}, \bibinfo{person}{Cliff
  Young}, \bibinfo{person}{Nishant Patil}, \bibinfo{person}{David Patterson},
  \bibinfo{person}{Gaurav Agrawal}, \bibinfo{person}{Raminder Bajwa},
  \bibinfo{person}{Sarah Bates}, \bibinfo{person}{Suresh Bhatia},
  \bibinfo{person}{Nan Boden}, \bibinfo{person}{Al Borchers},
  \bibinfo{person}{Rick Boyle}, \bibinfo{person}{Pierre-luc Cantin},
  \bibinfo{person}{Clifford Chao}, \bibinfo{person}{Chris Clark},
  \bibinfo{person}{Jeremy Coriell}, \bibinfo{person}{Mike Daley},
  \bibinfo{person}{Matt Dau}, \bibinfo{person}{Jeffrey Dean},
  \bibinfo{person}{Ben Gelb}, \bibinfo{person}{Tara~Vazir Ghaemmaghami},
  \bibinfo{person}{Rajendra Gottipati}, \bibinfo{person}{William Gulland},
  \bibinfo{person}{Robert Hagmann}, \bibinfo{person}{C.~Richard Ho},
  \bibinfo{person}{Doug Hogberg}, \bibinfo{person}{John Hu},
  \bibinfo{person}{Robert Hundt}, \bibinfo{person}{Dan Hurt},
  \bibinfo{person}{Julian Ibarz}, \bibinfo{person}{Aaron Jaffey},
  \bibinfo{person}{Alek Jaworski}, \bibinfo{person}{Alexander Kaplan},
  \bibinfo{person}{Harshit Khaitan}, \bibinfo{person}{Daniel Killebrew},
  \bibinfo{person}{Andy Koch}, \bibinfo{person}{Naveen Kumar},
  \bibinfo{person}{Steve Lacy}, \bibinfo{person}{James Laudon},
  \bibinfo{person}{James Law}, \bibinfo{person}{Diemthu Le},
  \bibinfo{person}{Chris Leary}, \bibinfo{person}{Zhuyuan Liu},
  \bibinfo{person}{Kyle Lucke}, \bibinfo{person}{Alan Lundin},
  \bibinfo{person}{Gordon MacKean}, \bibinfo{person}{Adriana Maggiore},
  \bibinfo{person}{Maire Mahony}, \bibinfo{person}{Kieran Miller},
  \bibinfo{person}{Rahul Nagarajan}, \bibinfo{person}{Ravi Narayanaswami},
  \bibinfo{person}{Ray Ni}, \bibinfo{person}{Kathy Nix},
  \bibinfo{person}{Thomas Norrie}, \bibinfo{person}{Mark Omernick},
  \bibinfo{person}{Narayana Penukonda}, \bibinfo{person}{Andy Phelps},
  \bibinfo{person}{Jonathan Ross}, \bibinfo{person}{Matt Ross},
  \bibinfo{person}{Amir Salek}, \bibinfo{person}{Emad Samadiani},
  \bibinfo{person}{Chris Severn}, \bibinfo{person}{Gregory Sizikov},
  \bibinfo{person}{Matthew Snelham}, \bibinfo{person}{Jed Souter},
  \bibinfo{person}{Dan Steinberg}, \bibinfo{person}{Andy Swing},
  \bibinfo{person}{Mercedes Tan}, \bibinfo{person}{Gregory Thorson},
  \bibinfo{person}{Bo Tian}, \bibinfo{person}{Horia Toma},
  \bibinfo{person}{Erick Tuttle}, \bibinfo{person}{Vijay Vasudevan},
  \bibinfo{person}{Richard Walter}, \bibinfo{person}{Walter Wang},
  \bibinfo{person}{Eric Wilcox}, {and} \bibinfo{person}{Doe~Hyun Yoon}.}
  \bibinfo{year}{2017}\natexlab{}.
\newblock \showarticletitle{In-Datacenter Performance Analysis of a Tensor
  Processing Unit}.
\newblock \bibinfo{journal}{\emph{SIGARCH Comput. Archit. News}}
  \bibinfo{volume}{45}, \bibinfo{number}{2} (\bibinfo{date}{June}
  \bibinfo{year}{2017}), \bibinfo{pages}{1–12}.
\newblock
\showISSN{0163-5964}
\urldef\tempurl%
\url{https://doi.org/10.1145/3140659.3080246}
\showDOI{\tempurl}


\bibitem[\protect\citeauthoryear{Kjolstad, Kamil, Chou, Lugato, and
  Amarasinghe}{Kjolstad et~al\mbox{.}}{2017}]%
        {taco}
\bibfield{author}{\bibinfo{person}{Fredrik Kjolstad}, \bibinfo{person}{Shoaib
  Kamil}, \bibinfo{person}{Stephen Chou}, \bibinfo{person}{David Lugato}, {and}
  \bibinfo{person}{Saman Amarasinghe}.} \bibinfo{year}{2017}\natexlab{}.
\newblock \showarticletitle{The Tensor Algebra Compiler}.
\newblock \bibinfo{journal}{\emph{Proc. ACM Program. Lang.}}
  \bibinfo{volume}{1}, \bibinfo{number}{OOPSLA}, Article
  \bibinfo{articleno}{77} (\bibinfo{date}{Oct.} \bibinfo{year}{2017}),
  \bibinfo{numpages}{29}~pages.
\newblock
\showISSN{2475-1421}
\urldef\tempurl%
\url{https://doi.org/10.1145/3133901}
\showDOI{\tempurl}


\bibitem[\protect\citeauthoryear{Krizhevsky, Sutskever, and Hinton}{Krizhevsky
  et~al\mbox{.}}{2012}]%
        {krizhevsky2012conv}
\bibfield{author}{\bibinfo{person}{Alex Krizhevsky}, \bibinfo{person}{Ilya
  Sutskever}, {and} \bibinfo{person}{Geoffrey~E. Hinton}.}
  \bibinfo{year}{2012}\natexlab{}.
\newblock \showarticletitle{Imagenet classification with deep convolutional
  neural networks}. In \bibinfo{booktitle}{\emph{Advances in Neural Information
  Processing Systems}}.
\newblock


\bibitem[\protect\citeauthoryear{Lattner, Amini, Bondhugula, Cohen, Davis,
  Pienaar, Riddle, Shpeisman, Vasilache, and Zinenko}{Lattner
  et~al\mbox{.}}{2020}]%
        {mlir}
\bibfield{author}{\bibinfo{person}{Chris Lattner}, \bibinfo{person}{Mehdi
  Amini}, \bibinfo{person}{Uday Bondhugula}, \bibinfo{person}{Albert Cohen},
  \bibinfo{person}{Andy Davis}, \bibinfo{person}{Jacques Pienaar},
  \bibinfo{person}{River Riddle}, \bibinfo{person}{Tatiana Shpeisman},
  \bibinfo{person}{Nicolas Vasilache}, {and} \bibinfo{person}{Oleksandr
  Zinenko}.} \bibinfo{year}{2020}\natexlab{}.
\newblock \bibinfo{title}{MLIR: A Compiler Infrastructure for the End of
  Moore's Law}.
\newblock
\newblock
\showeprint[arxiv]{2002.11054}~[cs.PL]


\bibitem[\protect\citeauthoryear{Markidis, Chien, Laure, Peng, and
  Vetter}{Markidis et~al\mbox{.}}{2018}]%
        {markidis2018tensorcore}
\bibfield{author}{\bibinfo{person}{Stefano Markidis}, \bibinfo{person}{Steven
  Wei~Der Chien}, \bibinfo{person}{Erwin Laure}, \bibinfo{person}{Ivy~Bo Peng},
  {and} \bibinfo{person}{Jeffrey~S. Vetter}.} \bibinfo{year}{2018}\natexlab{}.
\newblock \showarticletitle{NVIDIA Tensor Core Programmability, Performance \&
  Precision}.
\newblock \bibinfo{journal}{\emph{2018 IEEE International Parallel and
  Distributed Processing Symposium Workshops (IPDPSW)}} (\bibinfo{date}{May}
  \bibinfo{year}{2018}).
\newblock
\showISBNx{9781538655559}
\urldef\tempurl%
\url{https://doi.org/10.1109/ipdpsw.2018.00091}
\showDOI{\tempurl}


\bibitem[\protect\citeauthoryear{{Moreau}, {Chen}, {Vega}, {Roesch}, {Zheng},
  {Yan}, {Fromm}, {Jiang}, {Ceze}, {Guestrin}, and {Krishnamurthy}}{{Moreau}
  et~al\mbox{.}}{2019}]%
        {moreau2018vta}
\bibfield{author}{\bibinfo{person}{T. {Moreau}}, \bibinfo{person}{T. {Chen}},
  \bibinfo{person}{L. {Vega}}, \bibinfo{person}{J. {Roesch}},
  \bibinfo{person}{L. {Zheng}}, \bibinfo{person}{E. {Yan}}, \bibinfo{person}{J.
  {Fromm}}, \bibinfo{person}{Z. {Jiang}}, \bibinfo{person}{L. {Ceze}},
  \bibinfo{person}{C. {Guestrin}}, {and} \bibinfo{person}{A. {Krishnamurthy}}.}
  \bibinfo{year}{2019}\natexlab{}.
\newblock \showarticletitle{A Hardware-Software Blueprint for Flexible Deep
  Learning Specialization}.
\newblock \bibinfo{journal}{\emph{IEEE Micro}} (\bibinfo{year}{2019}),
  \bibinfo{pages}{1--1}.
\newblock
\showISSN{0272-1732}
\urldef\tempurl%
\url{https://doi.org/10.1109/MM.2019.2928962}
\showDOI{\tempurl}


\bibitem[\protect\citeauthoryear{Nandi, Willsey, Anderson, Wilcox, Darulova,
  Grossman, and Tatlock}{Nandi et~al\mbox{.}}{2020}]%
        {szalinski}
\bibfield{author}{\bibinfo{person}{Chandrakana Nandi}, \bibinfo{person}{Max
  Willsey}, \bibinfo{person}{Adam Anderson}, \bibinfo{person}{James~R. Wilcox},
  \bibinfo{person}{Eva Darulova}, \bibinfo{person}{Dan Grossman}, {and}
  \bibinfo{person}{Zachary Tatlock}.} \bibinfo{year}{2020}\natexlab{}.
\newblock \showarticletitle{Synthesizing Structured CAD Models with Equality
  Saturation and Inverse Transformations}. In
  \bibinfo{booktitle}{\emph{Proceedings of the 41st ACM SIGPLAN Conference on
  Programming Language Design and Implementation}} (London, UK)
  \emph{(\bibinfo{series}{PLDI 2020})}. \bibinfo{publisher}{Association for
  Computing Machinery}, \bibinfo{address}{New York, NY, USA},
  \bibinfo{pages}{31–44}.
\newblock
\showISBNx{9781450376136}
\urldef\tempurl%
\url{https://doi.org/10.1145/3385412.3386012}
\showDOI{\tempurl}


\bibitem[\protect\citeauthoryear{Newcomb, Adams, Johnson, Bodik, and
  Kamil}{Newcomb et~al\mbox{.}}{2020}]%
        {newcomb2020halide-rewrite}
\bibfield{author}{\bibinfo{person}{Julie~L. Newcomb}, \bibinfo{person}{Andrew
  Adams}, \bibinfo{person}{Steven Johnson}, \bibinfo{person}{Rastislav Bodik},
  {and} \bibinfo{person}{Shoaib Kamil}.} \bibinfo{year}{2020}\natexlab{}.
\newblock \showarticletitle{Verifying and Improving Halide’s Term Rewriting
  System with Program Synthesis}.
\newblock \bibinfo{journal}{\emph{Proc. ACM Program. Lang.}}
  \bibinfo{volume}{4}, \bibinfo{number}{OOPSLA}, Article
  \bibinfo{articleno}{166} (\bibinfo{date}{Nov.} \bibinfo{year}{2020}),
  \bibinfo{numpages}{28}~pages.
\newblock
\urldef\tempurl%
\url{https://doi.org/10.1145/3428234}
\showDOI{\tempurl}


\bibitem[\protect\citeauthoryear{Nvidia}{Nvidia}{2018}]%
        {nvdla}
\bibfield{author}{\bibinfo{person}{Nvidia}.} \bibinfo{year}{2018}\natexlab{}.
\newblock \bibinfo{title}{The NVIDIA Deep Learning Accelerator (NVDLA)}.
\newblock \bibinfo{howpublished}{\url{http://nvdla.org/}}.
\newblock


\bibitem[\protect\citeauthoryear{NVIDIA}{NVIDIA}{2020}]%
        {nvidia2020nhwc}
\bibfield{author}{\bibinfo{person}{NVIDIA}.} \bibinfo{year}{2020}\natexlab{}.
\newblock \bibinfo{title}{Convolutional Layers User Guide}.
\newblock
  \bibinfo{howpublished}{\url{https://docs.nvidia.com/deeplearning/performance/dl-performance-convolutional/index.html}}.
\newblock


\bibitem[\protect\citeauthoryear{Panchekha, Sanchez-Stern, Wilcox, and
  Tatlock}{Panchekha et~al\mbox{.}}{2015}]%
        {herbie}
\bibfield{author}{\bibinfo{person}{Pavel Panchekha}, \bibinfo{person}{Alex
  Sanchez-Stern}, \bibinfo{person}{James~R. Wilcox}, {and}
  \bibinfo{person}{Zachary Tatlock}.} \bibinfo{year}{2015}\natexlab{}.
\newblock \showarticletitle{Automatically Improving Accuracy for Floating Point
  Expressions}.
\newblock \bibinfo{journal}{\emph{SIGPLAN Not.}} \bibinfo{volume}{50},
  \bibinfo{number}{6} (\bibinfo{date}{June} \bibinfo{year}{2015}),
  \bibinfo{pages}{1–11}.
\newblock
\showISSN{0362-1340}
\urldef\tempurl%
\url{https://doi.org/10.1145/2813885.2737959}
\showDOI{\tempurl}


\bibitem[\protect\citeauthoryear{Paszke, Gross, Massa, Lerer, Bradbury, Chanan,
  Killeen, Lin, Gimelshein, Antiga, Desmaison, Köpf, Yang, DeVito, Raison,
  Tejani, Chilamkurthy, Steiner, Fang, Bai, and Chintala}{Paszke
  et~al\mbox{.}}{2019}]%
        {pytorch}
\bibfield{author}{\bibinfo{person}{Adam Paszke}, \bibinfo{person}{Sam Gross},
  \bibinfo{person}{Francisco Massa}, \bibinfo{person}{Adam Lerer},
  \bibinfo{person}{James Bradbury}, \bibinfo{person}{Gregory Chanan},
  \bibinfo{person}{Trevor Killeen}, \bibinfo{person}{Zeming Lin},
  \bibinfo{person}{Natalia Gimelshein}, \bibinfo{person}{Luca Antiga},
  \bibinfo{person}{Alban Desmaison}, \bibinfo{person}{Andreas Köpf},
  \bibinfo{person}{Edward Yang}, \bibinfo{person}{Zach DeVito},
  \bibinfo{person}{Martin Raison}, \bibinfo{person}{Alykhan Tejani},
  \bibinfo{person}{Sasank Chilamkurthy}, \bibinfo{person}{Benoit Steiner},
  \bibinfo{person}{Lu Fang}, \bibinfo{person}{Junjie Bai}, {and}
  \bibinfo{person}{Soumith Chintala}.} \bibinfo{year}{2019}\natexlab{}.
\newblock \bibinfo{title}{PyTorch: An Imperative Style, High-Performance Deep
  Learning Library}.
\newblock
\newblock
\showeprint[arxiv]{1912.01703}~[cs.LG]
\urldef\tempurl%
\url{https://arxiv.org/abs/1912.01703}
\showURL{%
\tempurl}


\bibitem[\protect\citeauthoryear{Qin, Klaaßen, Gallersdörfer, Stoll, and
  Zhang}{Qin et~al\mbox{.}}{2020}]%
        {qin2020bitcoins}
\bibfield{author}{\bibinfo{person}{Shize Qin}, \bibinfo{person}{Lena Klaaßen},
  \bibinfo{person}{Ulrich Gallersdörfer}, \bibinfo{person}{Christian Stoll},
  {and} \bibinfo{person}{Da Zhang}.} \bibinfo{year}{2020}\natexlab{}.
\newblock \bibinfo{title}{Bitcoin's future carbon footprint}.
\newblock
\newblock
\showeprint[arxiv]{2011.02612}~[econ.GN]


\bibitem[\protect\citeauthoryear{Ragan-Kelley, Barnes, Adams, Paris, Durand,
  and Amarasinghe}{Ragan-Kelley et~al\mbox{.}}{2013a}]%
        {halide}
\bibfield{author}{\bibinfo{person}{Jonathan Ragan-Kelley},
  \bibinfo{person}{Connelly Barnes}, \bibinfo{person}{Andrew Adams},
  \bibinfo{person}{Sylvain Paris}, \bibinfo{person}{Fr{\'e}do Durand}, {and}
  \bibinfo{person}{Saman Amarasinghe}.} \bibinfo{year}{2013}\natexlab{a}.
\newblock \showarticletitle{Halide: A Language and Compiler for Optimizing
  Parallelism, Locality, and Recomputation in Image Processing Pipelines}. In
  \bibinfo{booktitle}{\emph{Proceedings of the 34th ACM SIGPLAN Conference on
  Programming Language Design and Implementation}} (Seattle, Washington, USA)
  \emph{(\bibinfo{series}{PLDI '13})}. \bibinfo{publisher}{ACM},
  \bibinfo{address}{New York, NY, USA}, \bibinfo{pages}{519--530}.
\newblock
\showISBNx{978-1-4503-2014-6}
\urldef\tempurl%
\url{https://doi.org/10.1145/2491956.2462176}
\showDOI{\tempurl}


\bibitem[\protect\citeauthoryear{Ragan-Kelley, Barnes, Adams, Paris, Durand,
  and Amarasinghe}{Ragan-Kelley et~al\mbox{.}}{2013b}]%
        {ragan2013halide}
\bibfield{author}{\bibinfo{person}{Jonathan Ragan-Kelley},
  \bibinfo{person}{Connelly Barnes}, \bibinfo{person}{Andrew Adams},
  \bibinfo{person}{Sylvain Paris}, \bibinfo{person}{Fr{\'e}do Durand}, {and}
  \bibinfo{person}{Saman Amarasinghe}.} \bibinfo{year}{2013}\natexlab{b}.
\newblock \showarticletitle{Halide: a language and compiler for optimizing
  parallelism, locality, and recomputation in image processing pipelines}.
\newblock \bibinfo{journal}{\emph{Acm Sigplan Notices}} \bibinfo{volume}{48},
  \bibinfo{number}{6} (\bibinfo{year}{2013}), \bibinfo{pages}{519--530}.
\newblock


\bibitem[\protect\citeauthoryear{Reuther, Michaleas, Jones, Gadepally, Samsi,
  and Kepner}{Reuther et~al\mbox{.}}{2019}]%
        {reuther2019survey}
\bibfield{author}{\bibinfo{person}{Albert Reuther}, \bibinfo{person}{Peter
  Michaleas}, \bibinfo{person}{Michael Jones}, \bibinfo{person}{Vijay
  Gadepally}, \bibinfo{person}{Siddharth Samsi}, {and} \bibinfo{person}{Jeremy
  Kepner}.} \bibinfo{year}{2019}\natexlab{}.
\newblock \showarticletitle{Survey and Benchmarking of Machine Learning
  Accelerators}.
\newblock \bibinfo{journal}{\emph{2019 IEEE High Performance Extreme Computing
  Conference (HPEC)}} (\bibinfo{date}{Sep} \bibinfo{year}{2019}).
\newblock
\showISBNx{9781728150208}
\urldef\tempurl%
\url{https://doi.org/10.1109/hpec.2019.8916327}
\showDOI{\tempurl}


\bibitem[\protect\citeauthoryear{Roesch, Lyubomirsky, Kirisame, Pollock, Weber,
  Jiang, Chen, Moreau, and Tatlock}{Roesch et~al\mbox{.}}{2019}]%
        {relay}
\bibfield{author}{\bibinfo{person}{Jared Roesch}, \bibinfo{person}{Steven
  Lyubomirsky}, \bibinfo{person}{Marisa Kirisame}, \bibinfo{person}{Josh
  Pollock}, \bibinfo{person}{Logan Weber}, \bibinfo{person}{Ziheng Jiang},
  \bibinfo{person}{Tianqi Chen}, \bibinfo{person}{Thierry Moreau}, {and}
  \bibinfo{person}{Zachary Tatlock}.} \bibinfo{year}{2019}\natexlab{}.
\newblock \showarticletitle{Relay: {A} High-Level {IR} for Deep Learning}.
\newblock \bibinfo{journal}{\emph{CoRR}}  \bibinfo{volume}{abs/1904.08368}
  (\bibinfo{year}{2019}).
\newblock
\showeprint[arxiv]{1904.08368}
\urldef\tempurl%
\url{http://arxiv.org/abs/1904.08368}
\showURL{%
\tempurl}


\bibitem[\protect\citeauthoryear{{Simbürger}, {Apel}, {Größlinger}, and
  {Lengauer}}{{Simbürger} et~al\mbox{.}}{2013}]%
        {polyhedral-survey}
\bibfield{author}{\bibinfo{person}{A. {Simbürger}}, \bibinfo{person}{S.
  {Apel}}, \bibinfo{person}{A. {Größlinger}}, {and} \bibinfo{person}{C.
  {Lengauer}}.} \bibinfo{year}{2013}\natexlab{}.
\newblock \showarticletitle{The potential of polyhedral optimization: An
  empirical study}. In \bibinfo{booktitle}{\emph{2013 28th IEEE/ACM
  International Conference on Automated Software Engineering (ASE)}}.
  \bibinfo{pages}{508--518}.
\newblock
\urldef\tempurl%
\url{https://doi.org/10.1109/ASE.2013.6693108}
\showDOI{\tempurl}


\bibitem[\protect\citeauthoryear{Steuwer, Remmelg, and Dubach}{Steuwer
  et~al\mbox{.}}{2017}]%
        {lift}
\bibfield{author}{\bibinfo{person}{Michel Steuwer}, \bibinfo{person}{Toomas
  Remmelg}, {and} \bibinfo{person}{Christophe Dubach}.}
  \bibinfo{year}{2017}\natexlab{}.
\newblock \showarticletitle{Lift: A Functional Data-Parallel IR for
  High-Performance GPU Code Generation}. In
  \bibinfo{booktitle}{\emph{Proceedings of the 2017 International Symposium on
  Code Generation and Optimization}} (Austin, USA) \emph{(\bibinfo{series}{CGO
  '17})}. \bibinfo{publisher}{IEEE Press}, \bibinfo{pages}{74–85}.
\newblock
\showISBNx{9781509049318}


\bibitem[\protect\citeauthoryear{Tate, Stepp, Tatlock, and Lerner}{Tate
  et~al\mbox{.}}{2009}]%
        {tate2009equality}
\bibfield{author}{\bibinfo{person}{Ross Tate}, \bibinfo{person}{Michael Stepp},
  \bibinfo{person}{Zachary Tatlock}, {and} \bibinfo{person}{Sorin Lerner}.}
  \bibinfo{year}{2009}\natexlab{}.
\newblock \showarticletitle{Equality Saturation: A New Approach to
  Optimization}. In \bibinfo{booktitle}{\emph{Proceedings of the 36th Annual
  ACM Symposium on Principles of Programming Languages}}
  \emph{(\bibinfo{series}{POPL '09})}. \bibinfo{pages}{264–276}.
\newblock
\showISBNx{9781605583792}
\urldef\tempurl%
\url{https://doi.org/10.1145/1480881.1480915}
\showDOI{\tempurl}


\bibitem[\protect\citeauthoryear{Tian, Guo, Li, Ren, and Kestor}{Tian
  et~al\mbox{.}}{2021}]%
        {tian2021highperformance}
\bibfield{author}{\bibinfo{person}{Ruiqin Tian}, \bibinfo{person}{Luanzheng
  Guo}, \bibinfo{person}{Jiajia Li}, \bibinfo{person}{Bin Ren}, {and}
  \bibinfo{person}{Gokcen Kestor}.} \bibinfo{year}{2021}\natexlab{}.
\newblock \bibinfo{title}{A High-Performance Sparse Tensor Algebra Compiler in
  Multi-Level IR}.
\newblock
\newblock
\showeprint[arxiv]{2102.05187}~[cs.DC]


\bibitem[\protect\citeauthoryear{VanHattum, Nigam, Lee, Bornholt, and
  Sampson}{VanHattum et~al\mbox{.}}{2021}]%
        {vanhattum2021vectorization}
\bibfield{author}{\bibinfo{person}{Alexa VanHattum}, \bibinfo{person}{Rachit
  Nigam}, \bibinfo{person}{Vincent~T Lee}, \bibinfo{person}{James Bornholt},
  {and} \bibinfo{person}{Adrian Sampson}.} \bibinfo{year}{2021}\natexlab{}.
\newblock \showarticletitle{Vectorization for Digital Signal Processors via
  Equality Saturation}.
\newblock  (\bibinfo{year}{2021}).
\newblock


\bibitem[\protect\citeauthoryear{Vasilache, Zinenko, Theodoridis, Goyal,
  DeVito, Moses, Verdoolaege, Adams, and Cohen}{Vasilache
  et~al\mbox{.}}{2018}]%
        {vasilache2018tensor}
\bibfield{author}{\bibinfo{person}{Nicolas Vasilache},
  \bibinfo{person}{Oleksandr Zinenko}, \bibinfo{person}{Theodoros Theodoridis},
  \bibinfo{person}{Priya Goyal}, \bibinfo{person}{Zachary DeVito},
  \bibinfo{person}{William~S Moses}, \bibinfo{person}{Sven Verdoolaege},
  \bibinfo{person}{Andrew Adams}, {and} \bibinfo{person}{Albert Cohen}.}
  \bibinfo{year}{2018}\natexlab{}.
\newblock \showarticletitle{Tensor comprehensions: Framework-agnostic
  high-performance machine learning abstractions}.
\newblock \bibinfo{journal}{\emph{arXiv preprint arXiv:1802.04730}}
  (\bibinfo{year}{2018}).
\newblock


\bibitem[\protect\citeauthoryear{Wang, Hutchison, Suciu, Howe, and Leang}{Wang
  et~al\mbox{.}}{2020}]%
        {wang2020spores}
\bibfield{author}{\bibinfo{person}{Yisu~Remy Wang}, \bibinfo{person}{Shana
  Hutchison}, \bibinfo{person}{Dan Suciu}, \bibinfo{person}{Bill Howe}, {and}
  \bibinfo{person}{Jonathan Leang}.} \bibinfo{year}{2020}\natexlab{}.
\newblock \showarticletitle{{SPORES:} Sum-Product Optimization via Relational
  Equality Saturation for Large Scale Linear Algebra}.
\newblock \bibinfo{journal}{\emph{Proc. {VLDB} Endow.}} \bibinfo{volume}{13},
  \bibinfo{number}{11} (\bibinfo{year}{2020}), \bibinfo{pages}{1919--1932}.
\newblock
\urldef\tempurl%
\url{http://www.vldb.org/pvldb/vol13/p1919-wang.pdf}
\showURL{%
\tempurl}


\bibitem[\protect\citeauthoryear{Whitfield and Soffa}{Whitfield and
  Soffa}{1997}]%
        {phase-ordering}
\bibfield{author}{\bibinfo{person}{Deborah~L. Whitfield} {and}
  \bibinfo{person}{Mary~Lou Soffa}.} \bibinfo{year}{1997}\natexlab{}.
\newblock \showarticletitle{An Approach for Exploring Code Improving
  Transformations}.
\newblock \bibinfo{journal}{\emph{ACM Trans. Program. Lang. Syst.}}
  \bibinfo{volume}{19}, \bibinfo{number}{6} (\bibinfo{date}{Nov.}
  \bibinfo{year}{1997}), \bibinfo{pages}{1053–1084}.
\newblock
\showISSN{0164-0925}
\urldef\tempurl%
\url{https://doi.org/10.1145/267959.267960}
\showDOI{\tempurl}


\bibitem[\protect\citeauthoryear{Willsey, Nandi, Wang, Flatt, Tatlock, and
  Panchekha}{Willsey et~al\mbox{.}}{2021}]%
        {willsey2021egg}
\bibfield{author}{\bibinfo{person}{Max Willsey}, \bibinfo{person}{Chandrakana
  Nandi}, \bibinfo{person}{Yisu~Remy Wang}, \bibinfo{person}{Oliver Flatt},
  \bibinfo{person}{Zachary Tatlock}, {and} \bibinfo{person}{Pavel Panchekha}.}
  \bibinfo{year}{2021}\natexlab{}.
\newblock \showarticletitle{egg: fast and extensible equality saturation}.
\newblock \bibinfo{journal}{\emph{Proceedings of the ACM on Programming
  Languages}} \bibinfo{volume}{5}, \bibinfo{number}{POPL}
  (\bibinfo{year}{2021}), \bibinfo{pages}{1--29}.
\newblock


\bibitem[\protect\citeauthoryear{Yang, Gao, Liu, Setter, Pu, Nayak, Bell, Cao,
  Ha, Raina, Kozyrakis, and Horowitz}{Yang et~al\mbox{.}}{2020}]%
        {yang2020interstellar}
\bibfield{author}{\bibinfo{person}{Xuan Yang}, \bibinfo{person}{Mingyu Gao},
  \bibinfo{person}{Qiaoyi Liu}, \bibinfo{person}{Jeff Setter},
  \bibinfo{person}{Jing Pu}, \bibinfo{person}{Ankita Nayak},
  \bibinfo{person}{Steven Bell}, \bibinfo{person}{Kaidi Cao},
  \bibinfo{person}{Heonjae Ha}, \bibinfo{person}{Priyanka Raina},
  \bibinfo{person}{Christos Kozyrakis}, {and} \bibinfo{person}{Mark Horowitz}.}
  \bibinfo{year}{2020}\natexlab{}.
\newblock \showarticletitle{Interstellar: Using Halide's Scheduling Language to
  Analyze DNN Accelerators}. In \bibinfo{booktitle}{\emph{Proceedings of the
  Twenty-Fifth International Conference on Architectural Support for
  Programming Languages and Operating Systems}} (Lausanne, Switzerland)
  \emph{(\bibinfo{series}{ASPLOS '20})}. \bibinfo{publisher}{Association for
  Computing Machinery}, \bibinfo{address}{New York, NY, USA},
  \bibinfo{pages}{369–383}.
\newblock
\showISBNx{9781450371025}
\urldef\tempurl%
\url{https://doi.org/10.1145/3373376.3378514}
\showDOI{\tempurl}


\bibitem[\protect\citeauthoryear{Yang, Phothilimtha, Wang, Willsey, Roy, and
  Pienaar}{Yang et~al\mbox{.}}{[n.d.]}]%
        {yang2021equality}
\bibfield{author}{\bibinfo{person}{Yichen Yang},
  \bibinfo{person}{Phitchaya~Mangpo Phothilimtha}, \bibinfo{person}{Yisu~Remy
  Wang}, \bibinfo{person}{Max Willsey}, \bibinfo{person}{Sudip Roy}, {and}
  \bibinfo{person}{Jacques Pienaar}.} \bibinfo{year}{[n.d.]}\natexlab{}.
\newblock \showarticletitle{Equality Saturation for Tensor Graph
  Superoptimization}.
\newblock \bibinfo{journal}{\emph{arXiv preprint arXiv:2101.01332}}
  (\bibinfo{year}{[n.\,d.]}).
\newblock


\bibitem[\protect\citeauthoryear{Zheng, Jia, Sun, Wu, Yu, Haj-Ali, Wang, Yang,
  Zhuo, Sen, Gonzalez, and Stoica}{Zheng et~al\mbox{.}}{2020}]%
        {zheng2020ansor}
\bibfield{author}{\bibinfo{person}{Lianmin Zheng}, \bibinfo{person}{Chengfan
  Jia}, \bibinfo{person}{Minmin Sun}, \bibinfo{person}{Zhao Wu},
  \bibinfo{person}{Cody~Hao Yu}, \bibinfo{person}{Ameer Haj-Ali},
  \bibinfo{person}{Yida Wang}, \bibinfo{person}{Jun Yang},
  \bibinfo{person}{Danyang Zhuo}, \bibinfo{person}{Koushik Sen},
  \bibinfo{person}{Joseph~E. Gonzalez}, {and} \bibinfo{person}{Ion Stoica}.}
  \bibinfo{year}{2020}\natexlab{}.
\newblock \showarticletitle{Ansor: Generating High-Performance Tensor Programs
  for Deep Learning}. In \bibinfo{booktitle}{\emph{14th {USENIX} Symposium on
  Operating Systems Design and Implementation ({OSDI} 20)}}.
  \bibinfo{publisher}{{USENIX} Association}, \bibinfo{address}{Banff, Canada},
  \bibinfo{pages}{863--879}.
\newblock
\showISBNx{978-1-939133-19-9}
\urldef\tempurl%
\url{https://www.usenix.org/conference/osdi20/presentation/zheng}
\showURL{%
\tempurl}


\bibitem[\protect\citeauthoryear{Zuboff}{Zuboff}{2018}]%
        {surveillance}
\bibfield{author}{\bibinfo{person}{Shoshana Zuboff}.}
  \bibinfo{year}{2018}\natexlab{}.
\newblock \bibinfo{booktitle}{\emph{The Age of Surveillance Capitalism: The
  Fight for a Human Future at the New Frontier of Power}
  (\bibinfo{edition}{1st} ed.)}.
\newblock
\showISBNx{1610395697}


\end{thebibliography}

\section*{Broader Impact Statement}

The ability to develop effective compiler support for
  specialized hardware accelerators in ML,
  and HPC more broadly,
  has generally been restricted to a handful of
  elite, well-resourced teams.
This restriction slows hardware development
  and creates barriers to entry for teams in
  less privileged environments to contribute to
  and help guide the development of the field.

We believe that the access pattern abstraction
  and \g's approach to term rewriting for
  improving compiler support for custom
  accelerators will help advance both
  near-term practical and longer-term principled
  approaches to building flexible compiler infrastructure.
In turn, we hope that this infrastructure will
  help contribute to a broader, more diverse, and
  more inclusive community of folks working
  together to build efficient technologies for social
  good.
  
Of course, all technology is political and it can
  be difficult to anticipate how future
  researchers and practitioners may apply \g.
While the most obvious consequence of more
  efficient hardware utilization is better
  performance for users and lower environmental
  impact via decreased power consumption,
  it is also possible that access patterns and \g
  would enable the rapid obsoleting of current
  hardware platforms and therefore contribute
  to harmful electronic waste.
This work could also stimulate
  demand for hardware customization by
  removing compiler development--related overheads and
  ultimately lead to higher negative
  environmental impact similar to the
  situation with respect to custom ASICs
  for bitcoin mining~\cite{qin2020bitcoins}.

Also,
  any improvement to ML efficiency or applicability
  may contribute to economic and privacy concerns
  arising from increased technology company monopolization
  as discussed in Zuboff's
  \textit{The Age of Surveillance Capitalism}~\citep{surveillance}.

\end{document}